\documentclass[aps,prd,onecolumn,groupedaddress,floatfix,longbibliography,superscriptaddress]{revtex4}

\usepackage{amsmath}
\usepackage{amssymb}
\usepackage{amsfonts}
\usepackage{bbold}
\usepackage{bm}
\usepackage{cancel}
\usepackage{times,float}
\usepackage{graphicx}
\graphicspath{{Figures/}}
\usepackage[usenames,dvipsnames,svgnames]{xcolor}
\usepackage{hyperref}
\hypersetup{colorlinks=true, linkcolor=Blue, citecolor=Blue,urlcolor=Blue}
\usepackage{multirow}
\usepackage{ulem}
\usepackage{color,epsfig}
\usepackage{slashed}
\usepackage{feynmf}
\usepackage{array}
\usepackage{subcaption}
\usepackage{soul}

\begin{document}
\sloppy 

\title{Nonperturbative magnetotransport from band geometry in Weyl semimetals}

\author{L. Medel Onofre}
\email{leonardo.medel@correo.nucleares.unam.mx}
\address{Instituto de Ciencias Nucleares, Universidad Nacional Aut\'{o}noma de M\'{e}xico, 04510 Ciudad de M\'{e}xico, M\'{e}xico}

\author{A. Mart\'{i}n-Ruiz}
\email{alberto.martin@nucleares.unam.mx}
\address{Instituto de Ciencias Nucleares, Universidad Nacional Aut\'{o}noma de M\'{e}xico, 04510 Ciudad de M\'{e}xico, M\'{e}xico}

\begin{abstract}
We develop a nonperturbative semiclassical theory of magnetotransport in Weyl semimetals, retaining the full magnetic-field dependence of the Fermi-surface conductivity in the presence of Berry curvature and orbital magnetic moment effects. We obtain closed-form expressions valid to all orders in the magnetic field within the semiclassical regime. We show that the exact continuum formulation exhibits an intrinsic infrared sensitivity associated with the singular behavior of the orbital magnetic moment, requiring a physically motivated regularization. While the full conductivity tensor reduces to the standard quadratic magnetoconductivity, we demonstrate that magnetic-field expansion and momentum integration do not commute, leading to nonanalytic contributions at the level of scalar transport coefficients. Our results identify a regime, relevant for low carrier densities or moderate magnetic fields, where magnetotransport becomes intrinsically nonperturbative and cannot be captured by conventional weak-field expansions.
\end{abstract}

\maketitle

\section{Introduction}

Topological phases of matter have reshaped our understanding of electronic systems by revealing that global geometric properties of Bloch states can lead to robust and quantized physical responses \cite{Hasan2010,Qi2011}. Among these systems, topological semimetals, and in particular Weyl semimetals, occupy a central place due to their gapless nature and the presence of topologically protected band crossings acting as monopoles of Berry curvature in momentum space \cite{Wan2011,Burkov2011,Armitage2018}. These materials exhibit a variety of unconventional phenomena, including surface Fermi arcs, chiral anomaly effects, and anomalous transport responses \cite{Xu2015,Lv2015,Huang2015}, which have attracted considerable attention both from a fundamental perspective and for potential applications in electronic, spintronic, and sensing devices \cite{Yan2017,Armitage2018}.

Magnetotransport constitutes one of the most direct probes of the geometric and topological properties of Weyl systems. In particular, the interplay between Berry curvature and external electromagnetic fields gives rise to a rich phenomenology, including anomalous Hall responses, negative longitudinal magnetoresistance, and nonlinear transport effects \cite{Xiao2010,Son2013,Burkov2014,PhysRevB.103.035102}. A powerful framework to describe these phenomena is provided by semiclassical chiral kinetic theory, where Berry curvature modifies both the equations of motion and the phase-space measure \cite{Xiao2005,Son2013,Stephanov2012}. Within this approach, a number of works have successfully characterized magnetotransport properties in the weak-field regime, where the magnetic field is treated perturbatively and the response is expanded in powers of $B$ \cite{Kim2014,Lundgren2014,Sodemann2015}. While the weak-field quadratic magnetoconductivity associated with the chiral anomaly is well established experimentally \cite{Xiong2015,Huang2015,Zhang2016}, several measurements report clear deviations from this behavior at moderate fields, including saturation and nonuniversal curvature \cite{Li2016,Arnold2016}. These features are typically described using phenomenological fitting forms rather than derived from a microscopic theory.

However, restricting the analysis to the weak-field limit obscures important aspects of the semiclassical dynamics. In particular, in systems with low carrier density or under moderate magnetic fields, the dimensionless parameter controlling the expansion is no longer small, and higher-order effects can become relevant \cite{Gao2015,Goerbig2011}. This is especially significant in Weyl semimetals, where the orbital magnetic moment introduces a strong momentum dependence that enhances the role of low-energy states \cite{Xiao2005,Sundaram1999}. In such regimes, a nonperturbative treatment of the magnetic-field dependence is required in order to capture the full structure of the transport response \cite{Burkov2014,Spivak2016}.

{
Recent developments have considerably extended semiclassical transport theory by incorporating higher-order geometric corrections and field-induced modifications of Bloch-electron dynamics \cite{PhysRevLett.112.166601,PhysRevB.95.165135,PhysRevB.91.214405,Chang_2008}. These works have provided a systematic description of magnetotransport within the weak-field regime and have revealed a variety of geometric contributions to both linear and nonlinear electronic responses. In parallel, the orbital magnetic moment has been recognized as a key ingredient of Berry-curvature-induced transport, producing sizable corrections to magnetoconductivities through magnetic-field-induced deformations of the Fermi surface and their interplay with Berry curvature \cite{PhysRevB.111.035138,cwlr-h39s}. Rather than extending the semiclassical equations of motion to higher orders in the electromagnetic fields, the present work follows a complementary strategy: we retain the conventional first-order Berry-curvature and orbital-magnetic-moment framework, while treating its magnetic-field dependence exactly. As we show below, this approach uncovers qualitative features that are absent in finite-order weak-field expansions, including an intrinsic infrared sensitivity associated with the singular behavior of the orbital magnetic moment and the noncommutativity between magnetic-field expansion and momentum integration.
}

{
Besides externally applied electromagnetic fields, Weyl semimetals may also host emergent chiral gauge fields generated by smooth lattice deformations, strain, or crystalline defects. These perturbations shift the Weyl nodes in momentum space, giving rise to an effective axial gauge potential, $\mathbf{A}_5$, which couples with opposite signs to quasiparticles of opposite chirality. Spatial and temporal variations of this field generate pseudo-magnetic and pseudo-electric fields, $\mathbf{B}_5=\nabla\times\mathbf{A}_5$ and $\mathbf{E}_5=-\partial_t\mathbf{A}_5$, respectively. Such emergent fields provide a condensed-matter realization of chiral gauge fields and have attracted considerable attention owing to their ability to induce anomaly-related transport phenomena without externally applied magnetic fields. Their microscopic origin and physical consequences have been extensively investigated in strained Weyl and Dirac semimetals, as well as within more general geometric descriptions based on elastic and torsional backgrounds \cite{PhysRevLett.115.177202,PhysRevX.6.041021,PhysRevB.107.144206,Medel2023,sharma2026linearresponsetilteddirac}. These developments further illustrate the versatility of semiclassical transport theory and provide additional motivation for studying magnetotransport beyond the conventional weak-field regime.
}

In this work, we develop a nonperturbative semiclassical theory of magnetotransport in Weyl semimetals, focusing on the Fermi-surface contribution to the conductivity. By retaining the full magnetic-field dependence of the phase-space factor, generalized velocity, and band energy, we derive closed-form expressions for both isotropic and anisotropic components of the conductivity \cite{Sundaram1999,Xiao2010}. {The formalism is formulated in terms of the effective chiral electromagnetic fields, $\mathbf{E}_\chi=\mathbf{E}+\chi\mathbf{E}_5$ and $\mathbf{B}_\chi=\mathbf{B}+\chi\mathbf{B}_5$, thereby providing a unified description of electromagnetic, axial, and mixed electromagnetic-axial configurations. Unlike higher-order semiclassical approaches based on weak-field expansions, our treatment fully exploits the magnetic-field dependence already contained in the conventional Berry-curvature and orbital-magnetic-moment framework, without introducing additional higher-order corrections to the equations of motion.} We show that the resulting formulation exhibits an intrinsic infrared sensitivity associated with the singular behavior of the orbital magnetic moment \cite{Chang1996,Xiao2005}, and that a consistent treatment requires the introduction of a physically motivated infrared cutoff determined by the validity of the semiclassical approximation \cite{Ashcroft1976}. {Throughout this work we restrict our attention to the geometric contribution to magnetotransport arising from the Berry curvature and orbital magnetic moment. The role of the Lorentz-force streaming term, together with the corresponding regime of validity of the present treatment, is discussed in Sec.~\ref{semiclassical_section} and Appendix~\ref{app:Lorentz}.}

Our results provide a unified framework that connects the conventional weak-field expansion with a regime where magnetotransport becomes intrinsically nonperturbative. While the full conductivity tensor reduces to the standard quadratic-in-field form in the appropriate limit \cite{Son2013,Burkov2014}, the exact expressions reveal a nontrivial interplay between magnetic-field expansion and momentum integration, leading to nonanalytic features at the level of scalar transport coefficients \cite{Spivak2016}. This highlights the role of infrared physics in Weyl systems and identifies a regime where geometric and orbital effects go beyond the conventional perturbative description, even when the magnetic field is parametrically small \cite{Xiao2010}.

The remainder of this paper is organized as follows. In Sec.~\ref{semiclassical_section} we introduce the semiclassical framework and derive the general expression for the electric current, separating it into intrinsic and Fermi-surface contributions within chiral kinetic theory. In Sec.~\ref{Model} we present the linearized Weyl model and summarize its band structure and geometric properties, including Berry curvature and orbital magnetic moment. In Sec.~\ref{current_response_section} we analyze the current response, first evaluating the intrinsic contribution in closed form, and then deriving the Fermi-surface conductivity retaining the full magnetic-field dependence. In particular, we obtain exact expressions for both isotropic and anisotropic components and discuss their infrared structure and regularization. We further compare the nonperturbative results with the conventional weak-field expansion and establish their consistency at the level of the conductivity tensor. The technical details are relegated to the appendices. Finally, in Sec.~\ref{conclusion} we summarize our results and discuss their physical implications and possible extensions.

\section{Semiclassical transport and current decomposition in Weyl systems} \label{semiclassical_section}

The transport properties of Weyl semimetals are governed by the interplay between band dispersion and the geometric structure of Bloch states. In particular, the presence of Berry curvature endows quasiparticles with anomalous velocities and modifies the phase-space measure, while the orbital magnetic moment introduces an explicit field dependence in the energy spectrum \cite{Xiao2010,Sundaram1999,Chang1996}. These ingredients can be consistently incorporated within the framework of chiral kinetic theory, which provides a semiclassical description valid in the regime where Landau quantization can be neglected \cite{Son2013,Stephanov2012,Medel2023}.

To describe the dynamics near each Weyl node, we label the electronic states by a composite index $\alpha = ( s , \chi )$, where $s = \pm 1$ denotes the band index and $\chi = \pm 1$ the chirality. External electromagnetic fields and emergent axial gauge fields generated by strain or lattice deformations are incorporated through
\begin{align}
\mathbf{E} _{\chi} = \mathbf{E} + \chi \mathbf{E} _{5} , \qquad \mathbf{B} _{\chi} = \mathbf{B} + \chi \mathbf{B} _{5} ,
\end{align}
which couple to opposite chiralities with opposite signs \cite{Cortijo2015,Pikulin2016,Ilan2020}. Within this framework, the semiclassical equations of motion take the  \cite{Xiao2010}
\begin{align}
\dot{\mathbf{r}} _{\alpha} &= \frac{1}{\hbar} \nabla _{\mathbf{k}} \mathcal{E} _{\alpha \mathbf{k}}- \dot{\mathbf{k}} _{\alpha} \times \boldsymbol{\Omega} _{\alpha \mathbf{k} } , \qquad \hbar \dot{\mathbf k} _{\alpha} = - e \mathbf{E} _{\chi} - e \, \dot{\mathbf{r}} _{\alpha} \times \mathbf{B} _{\chi} , \label{eom2_combined}
\end{align}
where the Berry curvature is defined as $\boldsymbol{\Omega} _{\alpha \mathbf{k}} = i \left< \nabla_{\mathbf k}u_{\alpha\mathbf k} \right| \times \left| \nabla_{\mathbf k}u_{\alpha\mathbf k} \right>$. The energy dispersion is $\mathcal{E} _{\alpha \mathbf{k}} = E _{\alpha\mathbf k} - \mathbf m _{\alpha \mathbf{k}}\cdot \mathbf{B}_{\chi}$, which includes a Zeeman-like correction due to the orbital magnetic moment $\mathbf m_{\alpha\mathbf k}$, defined by $\mathbf{m} _{\alpha \mathbf{k}} = - i \, \frac{e}{2\hbar} \left< \nabla _{\mathbf{k}} u _{\alpha \mathbf{k}} \right| \times \bigl( \hat{H} - E _{\alpha \mathbf{k} } \bigr) \left| \nabla _{\mathbf{k}} u _{\alpha \mathbf{k}} \right>$. The Bloch eigenstates $\left| u _{\alpha \mathbf{k}} \right> $ satisfy the eigenvalue problem $\hat{H}_{\mathbf{k}} \left| u _{\alpha \mathbf{k}} \right>  = E _{\alpha\mathbf k}  \left| u _{\alpha \mathbf{k}} \right>  $, evaluated in the absence of magnetic fields, $B_{\chi}=0$.

Both the Berry curvature and the orbital magnetic moment introduce distinct field-dependent corrections to the semiclassical dynamics, which have direct consequences for magnetotransport. The Berry curvature gives rise to anomalous velocities and chiral effects, such as the anomalous Hall response and the chiral anomaly, while the orbital magnetic moment modifies the band energy and velocity, contributing to nonlinear transport and magnetic-field-induced anisotropies \cite{Son2013,Burkov2014,Xiao2005}. Together, these geometric quantities play a central role in shaping the magnetotransport response of Weyl systems \cite{Medel2023}.

Electric current is given by
\begin{align}
\mathbf{J} = - e \sum _{\alpha} \int \frac{d ^{3} \mathbf{k} }{ ( 2 \pi ) ^{3} } \; D _{\alpha \mathbf{k}} ^{-1} \; \dot{\mathbf{r}} _{\alpha} \, f _{\alpha \mathbf{k}}  ,
\end{align}
where $D _{\alpha \mathbf{k}} = \left( 1 + \frac{e}{\hbar} \mathbf{B} _{\chi} \cdot \boldsymbol{\Omega} _{\alpha \mathbf{k}} \right) ^{-1}$ is the Berry-curvature correction to the phase-space measure and $f _{\alpha \mathbf{k}} $ is the nonequilibrium distribution function, whose evolution is governed by the Boltzmann equation \cite{Xiao2010,Sundaram1999}
\begin{align}
\left( \frac{\partial}{\partial t} + \dot{\mathbf{r}} _{\alpha} \cdot \nabla _{\mathbf{r}} + \dot{\mathbf{k}} _{\alpha} \cdot \nabla _{\mathbf{k}} \right) f _{\alpha \mathbf{k}} = - \frac{ f _{\alpha \mathbf{k}} - f _{0} (\mathcal{E} _{\alpha \mathbf{k}} ) }{\tau},
\end{align}
where $f _{0} (\varepsilon) = [ 1 + e ^{\beta ( \varepsilon - \mu ) } ] ^{-1} $ is the Fermi-Dirac distribution and $\tau$ is the relaxation time.

For homogeneous stationary systems, this reduces to
\begin{align}
- \tau \, \dot{\mathbf{k}} _{\alpha} \cdot \nabla _{\mathbf{k}} f _{\alpha \mathbf{k}} = f _{\alpha \mathbf{k}} - f _{0} (\mathcal{E} _{\alpha \mathbf{k}} ) ,
\end{align}
whose leading-order solution in the electric field reads \cite{Son2013,Stephanov2012}
\begin{align}
f _{\alpha \mathbf{k}} = f _{0} (\mathcal{E} _{\alpha \mathbf{k}} ) + e \tau D _{\alpha \mathbf{k}}  \; \mathbf{E} _{\chi} \cdot \boldsymbol{\mathcal V} _{\alpha \mathbf{k}} \; \frac{\partial f _{0} (\mathcal{E} _{\alpha \mathbf{k}} ) }{\partial \mathcal{E} _{\alpha \mathbf{k}} }. \label{dist_function}
\end{align}
where we introduce the generalized velocity
\begin{align}
\boldsymbol{\mathcal V} _{\alpha \mathbf{k}} = \mathbf{v} _{\alpha \mathbf{k}} + \frac{e}{\hbar} ( \mathbf{v} _{\alpha \mathbf{k}} \cdot \boldsymbol{\Omega} _{\alpha \mathbf{k}}) \mathbf{B} _{\chi} , \label{generalized_velocity}
\end{align}
which incorporates the Berry-curvature correction to the semiclassical dynamics \cite{Xiao2010,Medel2023}. {Equation~(\ref{dist_function}) follows from retaining the electric-field contribution to the nonequilibrium distribution function while neglecting the explicit action of the magnetic Lorentz-force streaming term. Within this approximation, the magnetic-field dependence enters exclusively through the geometric ingredients of the semiclassical theory, namely the Berry-curvature-modified phase-space factor, the generalized velocity, and the orbital-magnetic-moment correction to the band energy. The Lorentz-force streaming term can be incorporated systematically within the same semiclassical framework, as discussed in detail in Appendix~\ref{app:Lorentz}.}

Substituting the semiclassical velocity (\ref{eom2_combined}) and the distribution function (\ref{dist_function}) into the current, it is convenient to reorganize the result in terms of physically distinct contributions. The total current separates into $\mathbf{J} = \mathbf{J}^{\mathrm{Hall}} + \mathbf{J}^{\mathrm{FS}}$. The intrinsic contribution arises from the anomalous velocity and is given by
\begin{align}
\mathbf{J}^{\mathrm{Hall}} = - \frac{e ^{2}}{\hbar} \, \sum _{\alpha} \mathbf{E} _{\chi} \times \int \frac{d ^{3} \mathbf{k} }{( 2 \pi ) ^{3} } \; \boldsymbol{\Omega} _{\alpha \mathbf{k} } \, f _{0}(\mathcal{E} _{\alpha\mathbf{k}} ) , \label{J_Hall}
\end{align}
which is independent of the relaxation time and therefore reflects purely geometric properties of the occupied states \cite{Xiao2010}.

The remaining contribution is associated with the nonequilibrium correction to the distribution function and defines the Fermi-surface current,
\begin{align}
\mathbf{J}^{\mathrm{FS}} = - e ^{2} \tau \sum _{\alpha} \int \frac{d ^{3} \mathbf{k} }{ ( 2 \pi ) ^{3} } \; D _{\alpha \mathbf{k} } \; \bigl( \mathbf{E} _{\chi} \cdot \boldsymbol{\mathcal{V}} _{\alpha \mathbf{k} } \bigr) \, \boldsymbol{\mathcal V} _{\alpha \mathbf{k} } \, \frac{\partial f _{0}(\mathcal{E}_{\alpha\mathbf k})}{\partial \mathcal{E}_{\alpha\mathbf k}}. \label{J_FS}
\end{align}
This term contains all dissipative contributions and depends explicitly on the scattering time. It incorporates both conventional transport and magnetic-field-induced corrections through the combined effect of the phase-space factor, the generalized velocity, and the field-dependent band energy \cite{Son2013,Burkov2014,Medel2023}.

{
The present analysis remains entirely within the conventional first-order semiclassical theory, including Berry-curvature and orbital-magnetic-moment corrections \cite{Xiao2010,Sundaram1999}. Its nonperturbative character stems from retaining the complete magnetic-field dependence generated by these first-order equations of motion, rather than performing a weak-field expansion or incorporating higher-order semiclassical corrections. Extensions of semiclassical dynamics beyond first order, including field-induced positional shifts and higher-order geometric effects, have been developed in Refs.~\cite{PhysRevLett.112.166601,PhysRevB.95.165135,PhysRevB.91.214405,Chang_2008}. The present work should therefore be viewed as a nonperturbative treatment of the conventional semiclassical framework, valid in the regime where Landau quantization can be neglected.

Within this framework, the inclusion of the Lorentz-force streaming term generates the conventional transverse Hall response and renormalizes the conductivity perpendicular to the magnetic field through the usual cyclotron factor, while leaving the strictly longitudinal conductivity unchanged. In particular, for the longitudinal configuration $\mathbf{E}_{\chi}\parallel\mathbf{B}_{\chi}$, or more generally in the weak-cyclotron regime, the conductivity expressions derived below are recovered. Throughout this work, we therefore focus on the geometric contribution to magnetotransport encoded in the Berry curvature and orbital magnetic moment, retaining their complete magnetic-field dependence. Accordingly, the conductivity expressions derived below are exact within the conventional first-order semiclassical framework.
}

In what follows, we evaluate these contributions retaining the full magnetic-field dependence of the semiclassical dynamics, which allows us to access the transport response beyond the weak-field regime.

\section{Linearized Weyl model: band structure and geometric properties} \label{Model}

We consider a minimal low-energy description of a time-reversal-breaking Weyl semimetal, consisting of two Weyl nodes of opposite chirality \cite{Wan2011,Burkov2011,Armitage2018}. Focusing on the vicinity of each node and neglecting nonuniversal corrections away from the crossing points, the Hamiltonian can be written as
\begin{align}
\hat{H} _{\mathbf{k}} = \chi \hbar v _{F}   \boldsymbol{\sigma} \cdot \mathbf{k} , \label{Hamiltonian}
\end{align}
where $v _{F}$ is the Fermi velocity, $\chi=\pm1$ labels the chirality, $\boldsymbol{\sigma}$ denotes the Pauli matrices, and $\mathbf{k}$ is the crystal momentum measured relative to the Weyl node. The energy spectrum follows immediately as
\begin{align}
E _{s \mathbf k} = s \hbar v _{F} k ,
\end{align}
where $s = \pm 1$ is the band index. The geometric properties of the bands are encoded in the Berry curvature, which for the Weyl Hamiltonian (\ref{Hamiltonian}) is \cite{Xiao2010}
\begin{align}
\boldsymbol{\Omega} _{\alpha \mathbf{k} } = - s \chi  \frac{\mathbf{k}}{2 k ^{3} } .
\end{align}
This result reflects the monopolar structure of the Berry curvature in momentum space, with each Weyl node acting as a source or sink of flux characterized by its chirality \cite{Armitage2018}. The orbital magnetic moment is found to be proportional to the Berry curvature. Indeed, one can spot the identity $\mathbf{m} _{\alpha \mathbf{k}} = (e/\hbar) \, E _{s \mathbf{k}} \, \boldsymbol{\Omega} _{\alpha \mathbf{k}}$ \cite{Chang1996,Xiao2005}.

\section{Current response and magnetotransport} \label{current_response_section}

In this section, we analyze the current response of Weyl fermions within the semiclassical framework introduced above, focusing on the effects of Berry curvature and orbital magnetic moment. Our aim is to develop a description of magnetotransport that remains valid beyond the weak-field regime by retaining the full dependence on the external magnetic field. To this end, we employ Eqs.~(\ref{J_Hall}) and (\ref{J_FS}), which describe the intrinsic Hall and Fermi-surface contributions to the current, respectively.

\subsection{Intrinsic contribution}

We now evaluate the intrinsic Hall-like current
\begin{align}
\mathbf{J}^{\mathrm{Hall}} =   \frac{e ^{2}}{2 \hbar} \,  \sum _{\alpha} s \chi \, \mathbf{E} _{\chi} \times \int \frac{d ^{3} \mathbf{k} }{ ( 2 \pi ) ^{3} } \,     \frac{ \hat{\mathbf{k}} }{k ^{2} } \; \Theta \left( \mu -  s \hbar v _{F} k - \frac{\chi e v _{F}}{2k} \, \hat{\mathbf{k}} \cdot \mathbf{B} _{\chi}  \right) , \label{J_intrinsic_exact}
\end{align}
where $\Theta(x)$ is the Heaviside step function. For Weyl nodes, the Berry curvature is radially symmetric, so that in the absence of a magnetic field the angular integral vanishes by symmetry. The orbital magnetic moment introduces an angular dependence through $\hat{\mathbf k} \cdot \mathbf{B} _{\chi}$, lifting this cancellation and yielding a finite intrinsic Hall response. For $\mu > 0$, the contribution arises solely from the conduction band; accordingly, we focus on this regime in the remainder of this work.

By rotational symmetry, the angular integrals are most conveniently evaluated in spherical coordinates. Since the integrand depends on $\hat{\mathbf k}$ only through $\hat{\mathbf k}\cdot \mathbf{B}_{\chi}$, the result is necessarily aligned with $\mathbf{B}_{\chi}$, while transverse contributions vanish upon angular averaging. Choosing the polar axis along $\hat{\mathbf{B}} _{\chi}$, Eq. (\ref{J_intrinsic_exact}) becomes
\begin{align}
\mathbf{J}^{\mathrm{Hall}} =   \frac{e ^{2}}{8 \pi ^{2} \hbar} \,  \sum _{\chi}  \chi \, \mathbf{E} _{\chi} \times \hat{\mathbf{B}} _{\chi} \int _{0} ^{\infty} dk \; \int _{-1} ^{1} d \xi \,  \xi  \; \Theta \left( \mu - \hbar v _{F} k - \frac{\chi e v _{F} B _{\chi} }{2k} \, \xi  \right) . \label{J_intrinsic_exact_2}
\end{align}
The remaining integral in Eq.~(\ref{J_intrinsic_exact_2}) can be evaluated exactly, yielding
\begin{align}
\mathbf{J}^{\mathrm{Hall}} =   \frac{e ^{2} k _{F} }{16 \pi h} \,  \sum _{\chi}   \mathbf{E} _{\chi} \times \hat{\mathbf{B}} _{\chi} \; \mathrm{sgn}(B _{\chi} ) \; \left[ \frac{ 1 - ( 1 + 6 | \lambda _{\chi} | ) ( 1 - 4 | \lambda _{\chi} | ) ^{3/2} }{15 | \lambda _{\chi} | ^{2} } + \frac{ 1 - ( 1 - 6 | \lambda _{\chi} | )( 1 + 4 | \lambda _{\chi} | ) ^{3/2} }{30 | \lambda _{\chi}| ^{2} } - 1 \right], \label{J_intrinsic_exact_FIN}
\end{align}
where $k _{F} = \frac{\mu}{\hbar v _{F}} $ is the Fermi momentum  and we have introduced the dimensionless parameter
\begin{align}
\lambda _{\chi} = \frac{ B _{\chi} }{B_{\mathrm{eff}}},
\qquad
B_{\mathrm{eff}} = \frac{2 \mu ^{2}}{e v _{F}^{2} \hbar}. \label{definition_lambda}
\end{align}
The detailed evaluation of the integral, including the treatment of both branches selected by the step function, is presented in Appendix~\ref{appendix_intrinsic}. The scale $B_{\mathrm{eff}}$ sets the characteristic magnetic field above which the semiclassical description begins to break down, i.e., where Landau quantization effects become relevant, and the parameter $\lambda_\chi$ therefore quantifies the strength of the magnetic field relative to this intrinsic scale \cite{Xiao2010,Goerbig2011}. The expression above provides a closed-form result for the intrinsic contribution, valid to all orders in the magnetic field within the semiclassical regime.

\begin{figure}
    \centering
    \includegraphics[width=0.6\linewidth]{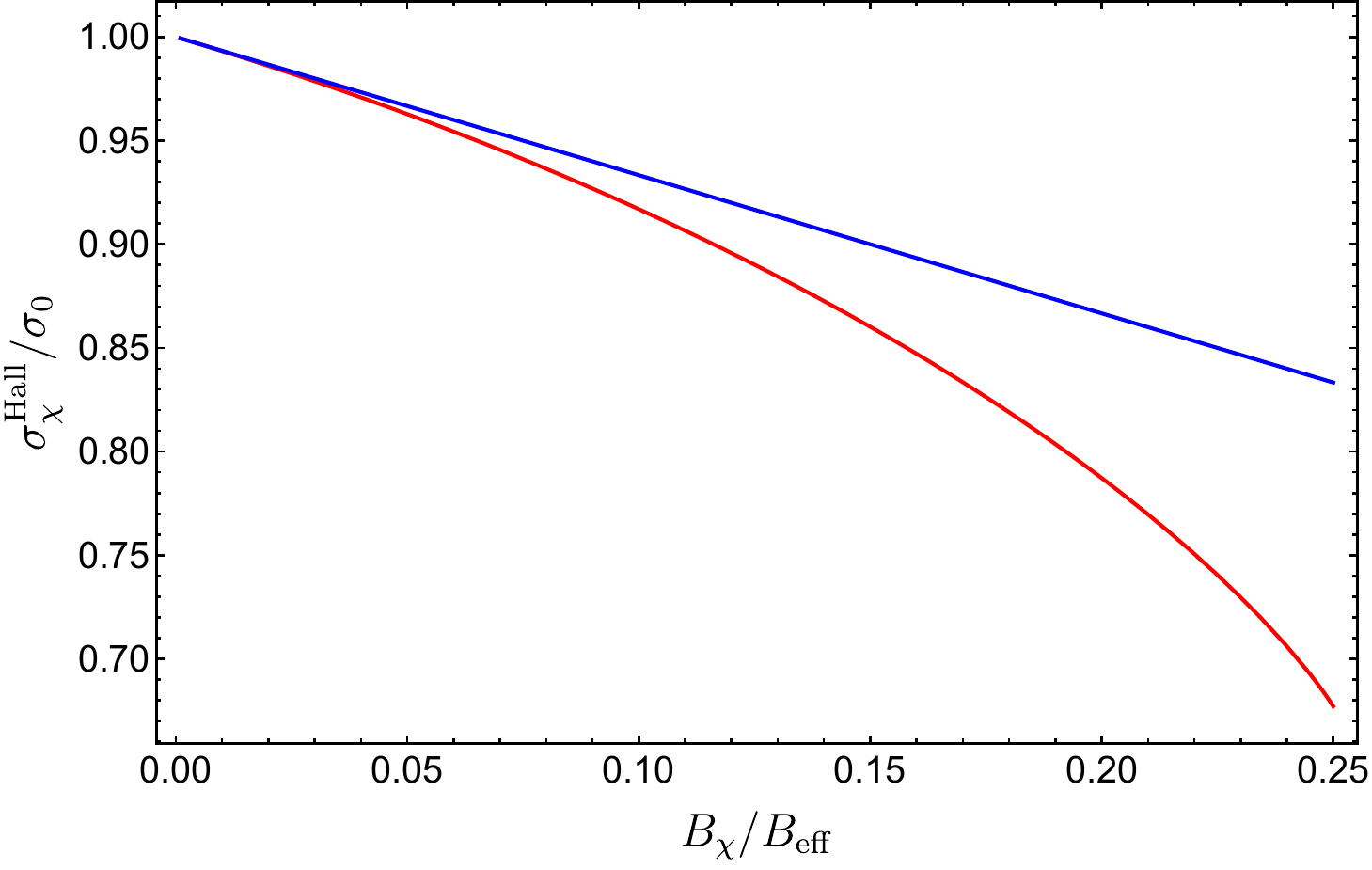}
    \caption{
Intrinsic Hall response conductivity $\sigma ^{\mathrm{Hall}} _{\chi} (B)$ (in units of $\sigma _{0} = \frac{e ^{2} k _{F} }{8 \pi h}$),  as a function of $\lambda_\chi = B_\chi/B_{\mathrm{eff}}$. The red curve corresponds to the exact nonperturbative result, Eq.~(\ref{J_intrinsic_exact_FIN}), while the blue curve shows the weak-field expansion. Both expressions agree in the limit $\lambda_\chi \to 0$, but deviations become apparent at moderate fields, where the exact response is suppressed relative to the linear approximation. This demonstrates the onset of nonperturbative effects beyond the weak-field regime.
}  \label{Hall_plot}
\end{figure}

To gain further insight into the magnetic-field dependence of the intrinsic contribution, in Fig.~\ref{Hall_plot} we plot the Hall scalar $\sigma^{\mathrm{Hall}}_{\chi}(B)$, defined by recasting the exact result in Eq.~(\ref{J_intrinsic_exact_FIN}) in the form
\begin{align}
\mathbf{J}^{\mathrm{Hall}}
=
\sum_{\chi}
\sigma^{\mathrm{Hall}}_{\chi}(B)\,
\mathbf{E}_{\chi}\times\hat{\mathbf{B}}_{\chi},
\label{J_Hall}
\end{align}
where $\lambda_\chi = B_\chi/B_{\mathrm{eff}}$.

{
Equation~(\ref{J_Hall}) explicitly identifies the intrinsic geometric Hall contribution arising from the anomalous velocity associated with the Berry curvature together with the orbital-magnetic-moment correction to the equilibrium occupation function within the conventional semiclassical theory. Since the homogeneous systems considered throughout this work do not support magnetization currents of the form $\mathbf{J}_{\rm mag}=\nabla\times\mathbf{M}$, the Hall current is entirely determined by the transport current.

The structure of Eq.~(\ref{J_Hall}) also clarifies how the contributions from different Weyl nodes combine. Because the current is expressed in terms of the effective chiral fields $\mathbf{E}_{\chi}=\mathbf{E}+\chi\mathbf{E}_{5}$ and $\mathbf{B}_{\chi}=\mathbf{B}+\chi\mathbf{B}_{5}$, the total Hall current is governed by the fields experienced by each chiral sector rather than by chirality alone. In the absence of axial fields, $\mathbf{E}_{5}=\mathbf{B}_{5}=0$, both nodes experience identical electromagnetic fields and therefore contribute constructively, yielding a total Hall response equal to twice the single-node result. More generally, strain-induced axial fields modify the relative contribution of each chirality through the combinations $\mathbf{E}+\chi\mathbf{E}_{5}$ and $\mathbf{B}+\chi\mathbf{B}_{5}$. Consequently, although particular field configurations may produce partial or complete cancellations, there is no general symmetry requiring the Hall-like contribution to vanish after summing over all Weyl nodes.
}

Having clarified the physical origin of the Hall contribution, we now compare the exact result with its weak-field expansion. As shown in Fig.~\ref{Hall_plot}, both curves coincide in the limit $\lambda_\chi\rightarrow0$, consistently reproducing the perturbative regime. However, as the magnetic field increases, the exact result deviates systematically from the linear approximation, exhibiting a stronger suppression. This behavior reflects the nonlinear dependence of the phase-space constraint induced by the orbital magnetic moment, which modifies the effective occupation of states away from the Fermi surface.

Importantly, the deviation becomes significant already for moderate values of $\lambda_\chi\sim0.1$, indicating that the weak-field expansion rapidly loses quantitative accuracy outside the strictly perturbative regime. This highlights the importance of retaining the complete magnetic-field dependence, particularly in systems with low carrier density, where the characteristic field scale $B_{\mathrm{eff}}$ is reduced and nonperturbative effects become experimentally accessible.

We now turn to the Fermi-surface contribution, which encodes the dissipative transport response and exhibits a richer structure due to the interplay between phase-space deformation, generalized velocity, and the field-dependent band energy.

\subsection{Fermi-surface contribution}

\begin{figure}[t]
\centering
\includegraphics[width=0.6\linewidth]{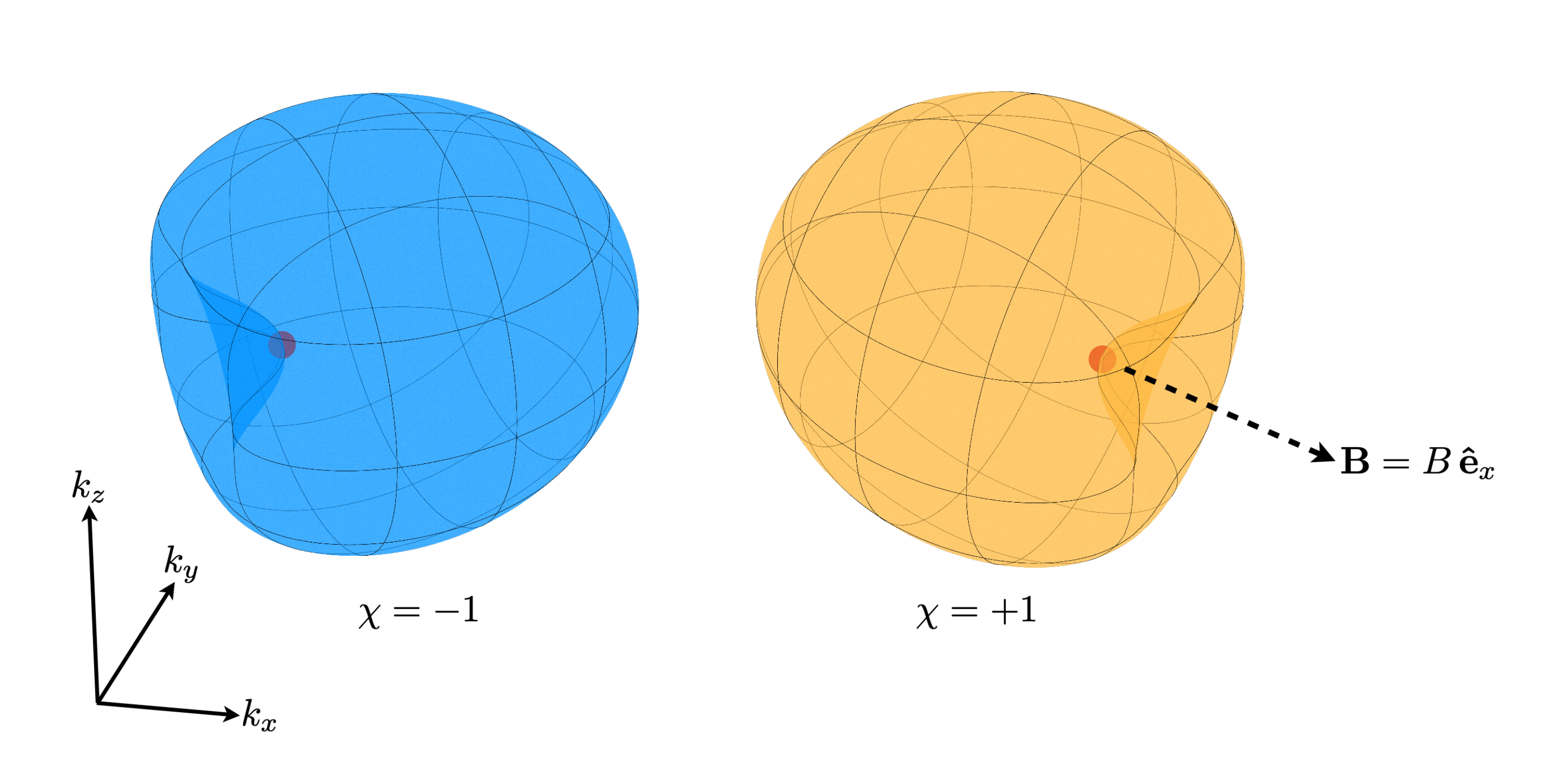}
\caption{ Fermi-surface deformation induced by the orbital magnetic moment in a Weyl semimetal.  The two panels correspond to opposite sectors labeled by $\chi=\pm1$.  In the absence of a magnetic field, the Fermi surface is isotropic around the axis defined by the magnetic field. The inclusion of the orbital magnetic moment introduces an explicit angular dependence through the term 
$\mathbf{m}_{\mathbf{k}}\cdot\mathbf{B}$, leading to a distortion of the constant-energy surface. 
As a result, the magnetic field acts as an angular selector, favoring specific regions of the Fermi surface 
(shaded regions), which dominate the transport response. This anisotropic redistribution of states 
underlies both the field-induced Hall response and the anisotropic Fermi-surface conductivity discussed in the text.
}
\label{fig:FS_OMM}
\end{figure}

We now turn to the Fermi-surface contribution to the current, given by \cite{Son2013,Kim2014}
\begin{align}
\mathbf{J} ^{\mathrm{FS}} =   e ^{2} \tau \sum _{\alpha} \int \frac{d ^{3} \mathbf{k} }{ ( 2 \pi ) ^{3} } \; D _{\alpha \mathbf{k} } \; \bigl( \mathbf{E} _{\chi} \cdot \boldsymbol{\mathcal{V}} _{\alpha \mathbf{k} } \bigr) \, \boldsymbol{\mathcal V} _{\alpha \mathbf{k} } \; \delta \left( \mu - s \hbar v _{F} k - \frac{\chi e v _{F}}{2k} \, \hat{\mathbf{k}} \cdot \mathbf{B} _{\chi}  \right) .  
\label{J_FS_2}
\end{align}
The constraint imposed by the Dirac delta function acquires a nontrivial angular dependence in the presence of the orbital magnetic moment. To gain physical insight into this constraint, it is useful to visualize the corresponding deformation of the Fermi surface. As shown in Fig.~\ref{fig:FS_OMM}, the orbital-magnetic-moment correction introduces an angular-dependent shift of the constant-energy surface, which depends explicitly on the projection $\hat{\mathbf{k}}\cdot \mathbf{B}_\chi$. In contrast to the zero-field case, where the Fermi surface is isotropic and contributions are uniformly distributed, the magnetic field selectively distorts different angular regions. As a result, the energy-conservation condition imposed by the delta function is satisfied only within specific sectors of momentum space.  This geometric selection mechanism provides a direct interpretation of the Fermi-surface contribution to the current. In particular, the anisotropy of the transport response originates from the uneven weighting of different regions of the distorted Fermi surface, while the chirality dependence reflects the opposite coupling of the orbital magnetic moment to the magnetic field. In this sense, magnetotransport is governed not only by dynamical quantities such as the velocity, but also by the magnetic-field-induced reshaping of the phase space available for conduction.

The term (\ref{J_FS_2}) originates from the nonequilibrium correction to the distribution function and is therefore proportional to the relaxation time $\tau$. Its evaluation requires the explicit form of the generalized velocity $\boldsymbol{\mathcal V}_{\alpha \mathbf{k}}$, defined by Eq.~(\ref{generalized_velocity}). In the Weyl semimetal model considered above, the generalized velocity becomes \cite{Xiao2010,Sundaram1999,Gao2015}
\begin{align}
\boldsymbol{\mathcal{V}} _{\alpha \mathbf{k}} = s v _{F} \hat{\mathbf{k}} - \chi \frac{e v _{F}}{\hbar} \frac{ \hat{\mathbf{k}} ( \hat{\mathbf{k}} \cdot \mathbf{B} _{\chi} ) }{k ^{2}}  + s \frac{e ^{2} v _{F}}{4 \hbar ^{2} } \frac{ \mathbf{B} _{\chi}  ( \hat{\mathbf{k}} \cdot \mathbf{B} _{\chi} ) }{k ^{4}} . 
\label{generalized_velocity_exact_final}
\end{align}

As before, we take $\mu > 0$, such that the contribution arises solely from the conduction band. The tensor structure of the angular integrals implies that the Fermi-surface current contains only an isotropic part proportional to $\mathbf{E} _{\chi}$ and a longitudinal anisotropic part proportional to $(\mathbf{E} _{\chi} \cdot \hat{\mathbf{B}} _{\chi} ) \hat{\mathbf{B}} _{\chi}$. Carrying out the angular integration, the Fermi-surface current can be decomposed as
\begin{align}
\mathbf{J} ^{\mathrm{FS}} = \sum _{\chi} \left[ \sigma _{\chi} (B) \, \mathbf{E} _{\chi} + \bar{\sigma} _{\chi} (B) \, ( \mathbf{E} _{\chi} \cdot \hat{\mathbf{B}} _{\chi} ) \, \hat{\mathbf{B}} _{\chi} \right] .
\label{J_FS_decomposition_main}
\end{align}
{It is worth emphasizing that Eq.~(\ref{J_FS_decomposition_main}) contains only isotropic and longitudinal anisotropic contributions. Within the present treatment, no transverse Hall term proportional to
$\hat{\mathbf B}_{\chi}\times\mathbf E_{\chi}$
appears, since the magnetic-field dependence enters exclusively through the Berry-curvature-modified phase-space factor, the generalized velocity, and the orbital-magnetic-moment correction to the band energy. The conventional Hall response associated with the explicit Lorentz-force streaming term in the Boltzmann equation is not included in Eq.~(\ref{J_FS_decomposition_main}). As discussed in Appendix~\ref{app:Lorentz}, retaining this term generates an additional transverse conductivity and renormalizes the conductivity perpendicular to the magnetic field, while leaving the longitudinal conductivity analyzed here unchanged. Consequently, the coefficients $\sigma_\chi(B)$ and $\bar{\sigma}_\chi(B)$ should be interpreted as the geometric contribution to the magnetotransport response within the conventional first-order semiclassical framework.

The symmetry considerations discussed after Eq.~(\ref{J_Hall}) apply equally to the Fermi-surface contribution in Eq.~(\ref{J_FS_decomposition_main}). In particular, the total response is obtained after summing over the chiral sectors and is therefore determined by the effective fields $\mathbf E_\chi$ and $\mathbf B_\chi$. Consequently, there is no generic cancellation of the Fermi-surface contribution upon summation over Weyl nodes.

}

In Eq.~(\ref{J_FS_decomposition_main}), $\sigma _{\chi} (B)$ denotes the isotropic longitudinal contribution, which already contains a magnetic-field dependence through the phase-space factor and the field-modified kinematics, while $\bar{\sigma} _{\chi} (B)$ captures the additional anisotropic response induced by the magnetic field. The isotropic coefficient is given by
\begin{align}
\sigma _{\chi} (B) = \frac{ e ^{2} v _{F}  \tau k _{F} ^{2}}{4 \pi h } \int _{0} ^{\infty} d \kappa \; \frac{\kappa ^{2} (3\kappa-2)^2}{2 \kappa - 1} \; \Theta \left(| \lambda _{\chi} |-\kappa | 1-\kappa | \right) \; \left[ \frac{1}{|\lambda _{\chi}|} - \frac{\kappa^2(1-\kappa)^2}{|\lambda _{\chi}|^3} \right]  , \label{sigma_iso_main}
\end{align}
whereas the anisotropic coefficient reads
\begin{align}
\bar{\sigma} _{\chi} (B) &= \frac{ e ^{2} v _{F} \tau k _{F} ^{2}}{4 \pi h } \int _{0} ^{\infty} d \kappa \; \frac{\kappa ^{2} }{2 \kappa - 1} \; \Theta \left(|\lambda _{\chi}|- \kappa | 1-\kappa | \right) \;  \Bigg\{ - \frac{(3\kappa-2)^2}{2|\lambda _{\chi}|} + \frac{3\kappa^2(1-\kappa)^2(3\kappa-2)^2}{2|\lambda _{\chi}|^3} \notag \\ & \hspace{8cm} + |\lambda _{\chi}|\frac{(1-\kappa)^2}{\kappa^4} + 2 \chi \,\mathrm{sgn}(\lambda _{\chi})\, \frac{(3\kappa-2)(1-\kappa)^2}{|\lambda _{\chi}|^2\,\kappa} \Bigg\}. \label{sigmaB_exact}
\end{align}

In the Appendix \ref{app:FS} we present a detailed derivation of these expressions. The conductivities given by Eqs.~(\ref{sigma_iso_main}) and (\ref{sigmaB_exact}) are therefore formally exact within the semiclassical framework. To evaluate these integrals, we assume $0<4|\lambda_\chi|<1$, which ensures that the system remains in the regime where Landau quantization does not set in and the semiclassical description is globally well defined \cite{Xiao2010,Goerbig2011}.  In this regime, the constraint imposed by the Heaviside function in Eqs.~(\ref{sigma_iso_main}) and (\ref{sigmaB_exact}) restricts the integration domain to the two intervals
\begin{align}
\kappa\in [0,\kappa_-]\cup[\kappa_+,\kappa_c],
\qquad
\kappa_\pm=\frac{1\pm\sqrt{1-4|\lambda_\chi|}}{2},
\qquad
\kappa_c=\frac{1+\sqrt{1+4|\lambda_\chi|}}{2}. \label{regions}
\end{align}
We now analyze the structure of both the isotropic and anisotropic coefficients. We start with the isotropic coefficient $\sigma_\chi(B)$ in Eq.~(\ref{sigma_iso_main}). In contrast to the anisotropic contribution in Eq. (\ref{sigmaB_exact}), the integrand entering $\sigma_\chi(B)$ is regular in the infrared limit $\kappa \to 0$, and therefore the exact expression can be evaluated without the need of introducing an infrared cutoff.

\begin{figure}
    \centering    \includegraphics[width=0.6\linewidth]{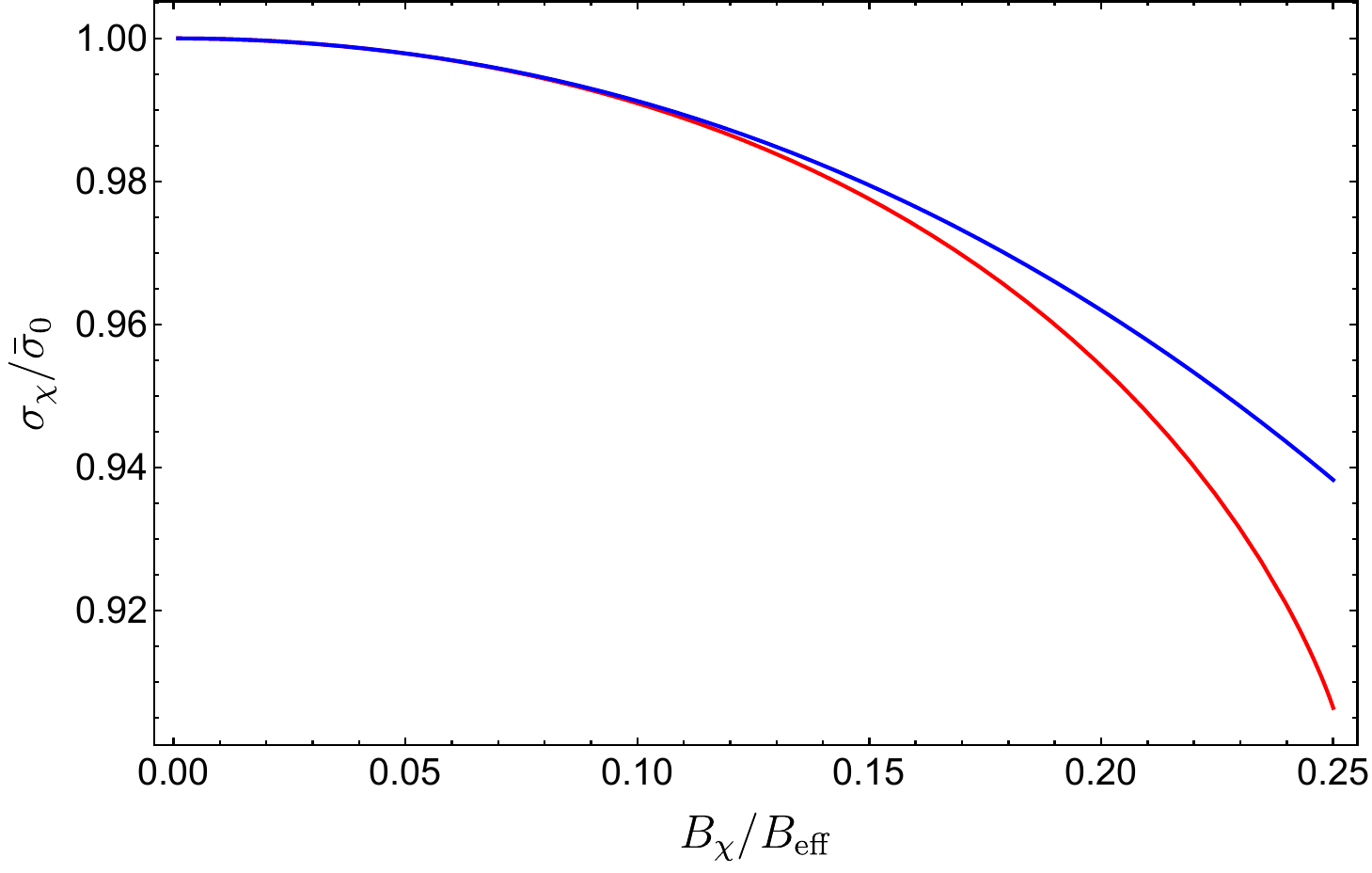}
    \caption{
Isotropic Fermi-surface conductivity $\sigma_\chi(B)$ (in units of $ \bar{\sigma} _{0} = \frac{e^2 v_F\tau k_F^2}{3 \pi h}$) as a function of $\lambda_\chi = B_\chi/B_{\mathrm{eff}}$. The red curve corresponds to the exact nonperturbative result, Eq.~(\ref{sigma_iso_exact}), while the blue curve shows the weak-field expansion. Both expressions agree in the limit $\lambda_\chi \to 0$, but deviations become visible at moderate magnetic fields, where the exact conductivity is suppressed relative to the perturbative approximation. In contrast to the anisotropic contribution, this coefficient is infrared safe, and the observed deviations reflect genuine nonperturbative effects beyond the weak-field regime.
} \label{isotropic_plot}
\end{figure}

Performing the integral in Eq.~(\ref{sigma_iso_main})  over these regions one obtains the closed-form expression
\begin{align}
\sigma_\chi(B)
=
\frac{e^2 v_F\tau k_F^2}{4\pi h}
\frac{1}{107520\,|\lambda_\chi|^3}
\Big[
&60480 |\lambda_\chi|^4 + 31360 |\lambda_\chi|^3 - 840 |\lambda_\chi|^2 + 420 |\lambda_\chi| + 256
\notag\\
&+\left(61440 |\lambda_\chi|^3 - 3072 |\lambda_\chi|^2 - 1024 |\lambda_\chi| - 512\right)\sqrt{1-4|\lambda_\chi|}
\notag\\
&+\left(30720 |\lambda_\chi|^3 + 1536 |\lambda_\chi|^2 - 512 |\lambda_\chi| + 256\right)\sqrt{1+4|\lambda_\chi|}
\notag\\
&+105\bigl(16|\lambda_\chi|^2-1\bigr)\ln(1+4|\lambda_\chi|)
\Big].
\label{sigma_iso_exact}
\end{align}
This result is fully nonperturbative in the magnetic-field parameter $\lambda_\chi$, and in contrast to the anisotropic coefficient, it is manifestly infrared safe. In particular, the exact expression remains finite without any additional regularization, reflecting the absence of singular contributions from the low-momentum region. Its weak-field expansion is therefore well defined directly at the scalar level and reads
\begin{align}
\sigma_\chi(B)
=
\frac{ e ^{2} v _{F} \tau k _{F} ^{2}}{4 \pi h }
\left[
\frac{4}{3}
-\frac{16}{15}|\lambda_\chi|^2
-|\lambda_\chi|^3
-\frac{118}{35}|\lambda_\chi|^4
+\mathcal O(|\lambda_\chi|^5)
\right].
\label{sigma_iso_weak}
\end{align} 
To further elucidate the structure of the Fermi-surface contribution, in Fig.~\ref{isotropic_plot} we plot the isotropic conductivity $\sigma_\chi(B)$, Eq.~(\ref{sigma_iso_exact}), in units of $\bar{\sigma} _{0} = \frac{e^2 v_F\tau k_F^2}{3 \pi h}$, as a function of the dimensionless parameter $\lambda_\chi = B_\chi/B_{\mathrm{eff}}$, and compare it with its weak-field expansion. As shown in the figure, both curves coincide in the limit $\lambda_\chi \to 0$, consistently reproducing the perturbative result. However, as the magnetic field increases, the exact expression deviates systematically from the weak-field approximation, exhibiting a stronger suppression of the conductivity. This behavior reflects the nonlinear dependence of the phase-space factor and the generalized velocity on the magnetic field, which modifies the effective transport properties beyond a simple quadratic correction. Since the isotropic coefficient is free from infrared singularities, the deviation observed in Fig.~\ref{isotropic_plot} should be interpreted as a genuine nonperturbative effect, rather than a manifestation of infrared sensitivity. Notably, the breakdown of the weak-field approximation occurs already at moderate values of $\lambda_\chi$, indicating that higher-order corrections become relevant well before reaching the regime where Landau quantization sets in. This highlights that even in the absence of infrared divergences, the full magnetic-field dependence encoded in the exact expression is essential for a quantitatively accurate description of magnetotransport, particularly in systems with low carrier density where the effective scale $B_{\mathrm{eff}}$ is reduced.

On the other hand, the structure of Eq.~(\ref{sigmaB_exact}) reveals that the anisotropic contribution $\bar{\sigma}_\chi(B)$ is particularly sensitive to the infrared sector. This behavior originates from the singular orbital magnetic moment, $\mathbf{m}_{\alpha\mathbf{k}}\sim 1/k$ \cite{Xiao2010,Gao2015}, which enhances the contribution of small momenta and signals that the exact continuum expression probes a region where the linearized Weyl Hamiltonian and the semiclassical expansion cease to be strictly reliable. For this reason, rather than taking the weak-field limit as the starting point of the analysis, it is more appropriate to first regularize the infrared behavior of the exact expression and only then extract the perturbative regime. 
This procedure is crucial, as it ensures that the magnetic-field dependence is extracted from a well-defined expression. In particular, it prevents spurious contributions associated with the unphysical extension of the continuum model into the deep infrared, where lattice effects are expected to become relevant.

The evaluation of Eq.~(\ref{sigmaB_exact}) must be performed in a way that remains consistent with the semiclassical description. The key point is that the continuum Weyl model ceases to be reliable at sufficiently small momenta, where the OMM correction to the energy becomes comparable to the band energy itself. Therefore, the validity of the semiclassical approximation requires that the OMM correction, $\frac{\chi e v_F}{2k}\,\hat{\mathbf k}\cdot \mathbf B_\chi$, remains small compared to the band energy $\hbar v_F k$, which implies
\begin{align}
\kappa^2 \gg |\lambda_\chi|.
\end{align}
{
Interestingly, this condition admits an alternative physical interpretation in terms of the conventional criterion for the validity of semiclassical dynamics. For Weyl fermions, the semiclassical regime requires that Landau quantization remain negligible, or equivalently, that many Landau levels participate in the dynamics. Using the Weyl Landau-level spectrum,
\[
E_n=v_F\sqrt{2e\hbar B_\chi n},
\]
together with the semiclassical relation $E=\hbar v_F k$, one finds that the condition $n\gg1$ translates into
\[
k^2 \gg \frac{eB_\chi}{\hbar},
\]
which coincides parametrically with the condition derived above from the orbital-magnetic-moment correction. Thus, the onset of the infrared regime identified by our semiclassical analysis occurs at the same characteristic momentum scale that marks the breakdown of the semiclassical approximation from the conventional Landau-level perspective.
}

The breakdown of the semiclassical description at momenta satisfying
\[
\kappa^2 \lesssim |\lambda_\chi|
\]
therefore requires the introduction of an infrared cutoff,
\begin{align}
\kappa_{\mathrm{IR}}
\sim
\sqrt{|\lambda_\chi|}.
\label{IR_cutoff_definition}
\end{align}
{
The infrared cutoff introduced in this way is therefore not an \textit{ad hoc} regularization prescription, but rather a direct manifestation of the standard criterion for the validity of the semiclassical approximation.
}

For $0<4|\lambda_\chi|<1 $ one finds $\kappa_{\mathrm{IR}}>\kappa_-$, so that the low-momentum branch in Eq. (\ref{regions}) is entirely removed by the infrared cutoff. As a consequence, the regularized exact expression receives contributions only from the interval
\begin{align}
\kappa\in[\kappa_+,\kappa_c].
\end{align}
This corresponds precisely to the portion of the Fermi surface that remains after excluding the infrared cap discussed above. Therefore, the anisotropic coefficient (\ref{sigmaB_exact}) can be written compactly as
\begin{align}
\bar{\sigma}_\chi(B)
=
\frac{e^2 v_F \tau k_F^2}{4\pi h}
\left[
\mathcal F_\chi(\kappa_c)-\mathcal F_\chi(\kappa_+ )
\right],
\label{sigmaB_exact_closed_compact}
\end{align}
where we defined
\begin{align}
\mathcal F_\chi(\kappa)
&=
\frac{|\lambda_\chi|}{\kappa}
-\frac{\kappa(\kappa-1)\bigl(18\kappa^2-2\kappa-1\bigr)}{32 |\lambda_\chi| }
+ \chi\,\mathrm{sgn}(\lambda_\chi) \,
\frac{\kappa\bigl(18\kappa^3-52\kappa^2+45\kappa-3\bigr)}{24 |\lambda_\chi|^2}
\notag\\
&\quad
+\frac{\kappa\bigl(15120\kappa^7-48960\kappa^6+54320\kappa^5-21168\kappa^4+210\kappa^3+140\kappa^2+105\kappa+105\bigr)}{17920\,|\lambda_\chi|^3}
\notag\\
&\quad
+\frac{512 |\lambda_\chi| ^4-16 |\lambda_\chi| ^2-64 |\lambda_\chi| \chi\,\mathrm{sgn}(\lambda_\chi) +3}{1024\,|\lambda_\chi|^3}\,
\ln|2\kappa-1| .
\label{primitive_sigmaB_compact}
\end{align}
Equation~(\ref{sigmaB_exact_closed_compact}) is nonperturbative in the magnetic field parameter $\lambda_\chi$, while remaining fully consistent with the semiclassical approximation through the infrared cutoff~(\ref{IR_cutoff_definition}). Its main physical content is twofold. First, it retains the full field dependence of the anisotropic response and therefore goes beyond the standard weak-field expansion. Second, it makes explicit that the exact continuum Weyl theory is infrared sensitive, and that a microscopic low-momentum regularization is required whenever one wishes to treat the orbital-moment contribution nonperturbatively.

We now analyze the weak-field regime, defined by $| \lambda _{\chi} | \ll 1$. In this limit, the anisotropic conductivity becomes
\begin{align}
\bar{\sigma}_\chi(B)
=
\frac{ e ^{2} v _{F} \tau k _{F} ^{2}}{4 \pi h }
\left[
\frac{4}{3}\,\chi\,\lambda_\chi
-\frac{8}{15}|\lambda_\chi|^2
+\frac{16}{5}\,\chi\,\lambda_\chi |\lambda_\chi|^2
-10|\lambda_\chi|^4
+\mathcal O(|\lambda_\chi|^5)
\right].
\label{sigmaB_weak_regulated}
\end{align}
{
Although this expression contains a term linear in $\lambda_\chi$, this contribution is nonanalytic in the magnetic field and therefore cannot be interpreted as part of a conventional weak-field expansion. The appearance of such terms reflects the nonperturbative nature of the exact regularized conductivity and the fact that magnetic-field expansion and momentum integration do not generally commute. Consequently, the weak-field behavior must be analyzed separately through a perturbative expansion performed prior to momentum integration, as discussed in Appendix~\ref{app:perturbative_tensor}.}

It is instructive to compare the weak-field limit of the nonperturbative expressions derived above with the standard perturbative expansion presented in Appendix~\ref{app:perturbative_tensor}. In the Appendix, the conductivity tensor is constructed by expanding systematically the phase-space factor, the generalized velocity, and the distribution function up to second order in the magnetic field. This procedure leads to the well-known result that the linear contribution in $B$ vanishes identically after angular integration, and the leading correction arises at order $B^{2}$. For completeness, the perturbative result for the conductivity tensor per chirality reads \cite{PhysRevB.94.245121, Medel2023, Medel2024}
\begin{align}
\sigma_{ij}^{(\chi)}
=
\frac{e^{2}v_F\tau k_F^{2}}{3\pi h}
\left[
\left(1-\frac{2}{5}\lambda_{\chi}^{2}\right)\delta_{ij}
+
\frac{6}{5}\lambda_{\chi}^{2}\,
\hat B_{\chi i}\hat B_{\chi j}
\right]
+
\mathcal O(\lambda_{\chi}^{4}),
\label{sigma_tensor_perturbative_main}
\end{align}
which provides a convenient benchmark for the weak-field behavior.

\begin{figure}
    \centering    \includegraphics[width=0.6\linewidth]{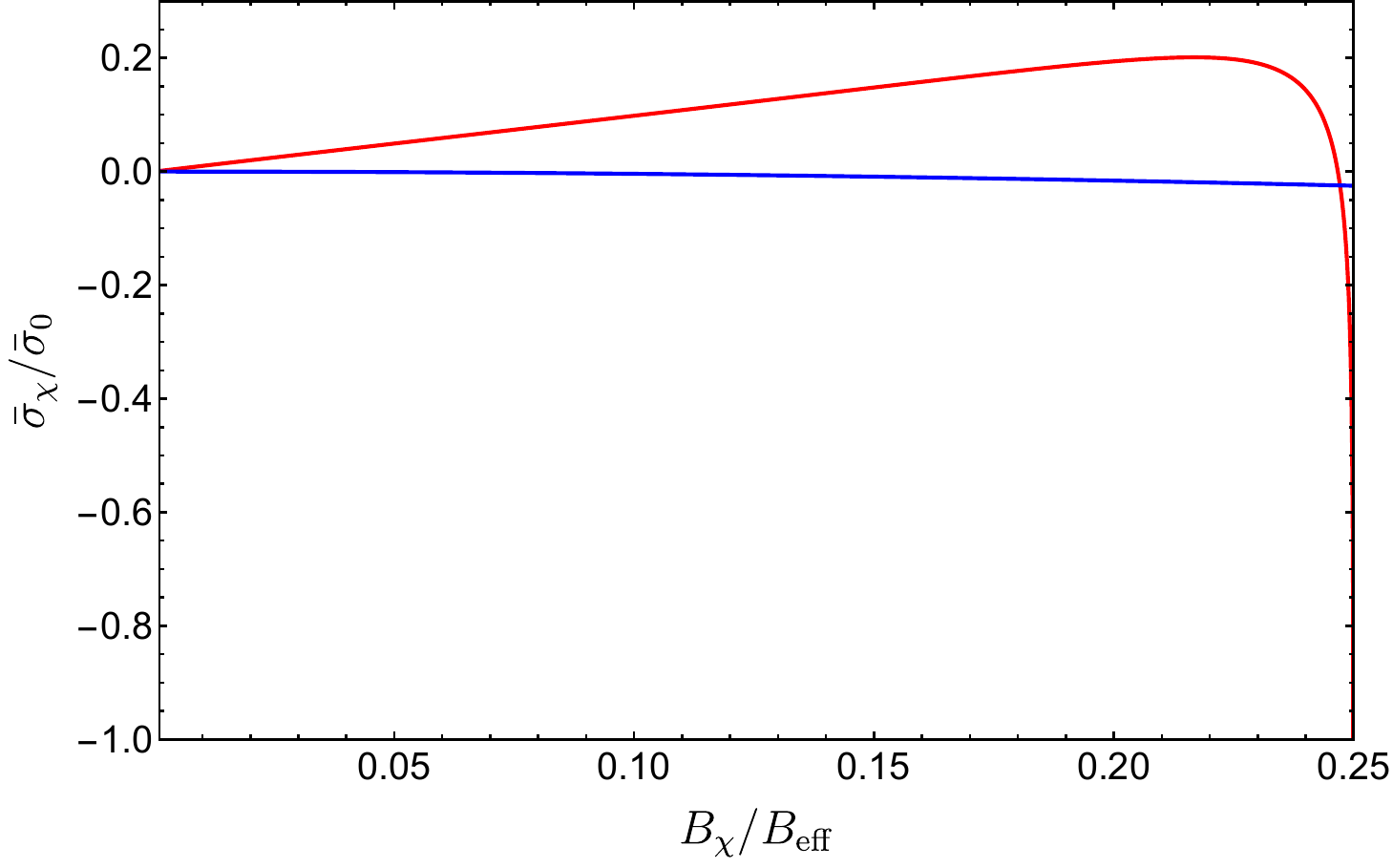}
    \caption{ Anisotropic Fermi-surface conductivity $\bar{\sigma}_\chi(B)$ (in units of $\bar{\sigma} _{0} = \frac{e^2 v_F\tau k_F^2}{3 \pi h}$) as a function of $\lambda_\chi = B_\chi/B_{\mathrm{eff}}$. The red curve corresponds to the exact nonperturbative result, Eq.~(\ref{sigmaB_exact_closed_compact}), while the blue curve shows the perturbative result (\ref{sigma_tensor_perturbative_main}). In contrast to the isotropic contribution, the two results do not coincide even in the weak-field limit, reflecting the intrinsic infrared sensitivity of the anisotropic response. The deviation originates from the singular behavior of the orbital magnetic moment and the resulting nonanalytic dependence on the magnetic field.} \label{anisotropic_plot}
\end{figure}

By contrast, the nonperturbative evaluation of the Fermi-surface integrals reveals a more intricate structure. In particular, the weak-field expansion of the anisotropic coefficient, Eq.~(\ref{sigmaB_weak_regulated}), contains a term proportional to $\chi\,\lambda_\chi$, which is formally linear in the magnetic field. However, this contribution is nonanalytic in $B$, as it originates from the infrared-sensitive sector of the exact expression and depends explicitly on the cutoff scale $\kappa_{\mathrm{IR}}\sim\sqrt{|\lambda_\chi|}$. As a consequence, this term does not correspond to a genuine perturbative contribution and must cancel out when reconstructing the full conductivity tensor. This explains why no linear-in-$B$ term appears in the perturbative treatment, where the expansion is performed at the level of the integrand before integration, and infrared singularities are never probed.

To complete the analysis of the Fermi-surface contribution, in Fig.~\ref{anisotropic_plot} we present the anisotropic conductivity $\bar{\sigma}_\chi(B)$, comparing the exact nonperturbative result obtained from Eq.~(\ref{sigmaB_exact_closed_compact}) with the perturbative expression derived in Appendix~\ref{app:perturbative_tensor}, which corresponds to the standard result reported in the literature. In contrast to the isotropic contribution, the agreement between both approaches is not recovered even in the weak-field limit. As shown in the figure, the exact result exhibits a systematic deviation from the perturbative prediction already at very small values of $\lambda_\chi$, indicating that the anisotropic response is intrinsically sensitive to nonperturbative effects. The origin of this discrepancy can be traced back to the infrared structure of the continuum Weyl model.

Importantly, this infrared regularization modifies the structure of the conductivity already at leading order, leading to a nonanalytic dependence on the magnetic field. As a consequence, the perturbative expansion, which assumes analyticity in $B$ and effectively excludes the infrared sector from the outset, fails to capture the correct behavior of the anisotropic response, even in the nominal weak-field regime. This demonstrates explicitly that the operations of magnetic-field expansion and momentum integration do not commute in the continuum Weyl model.

From a physical perspective, the deviations observed in Fig.~\ref{anisotropic_plot} indicate that the anisotropic magnetotransport response is governed by an interplay between Berry curvature, orbital magnetic moment, and infrared physics, and therefore cannot be accurately described within a purely perturbative framework. This suggests that experimental measurements of anisotropic magnetoconductivity may provide a direct probe of these nonperturbative effects, especially in systems with low carrier density where the infrared sector becomes more prominent.

On the other hand, the comparison at zeroth and second order in the magnetic field shows complete agreement between both approaches. From Eq.~(\ref{sigma_iso_weak}), the isotropic conductivity admits the expansion
\begin{align}
\sigma_\chi(B)
=
\frac{ e ^{2} v _{F} \tau k _{F} ^{2}}{4 \pi h }
\left[
\frac{4}{3}
-\frac{16}{15}|\lambda_\chi|^2
+\mathcal O(|\lambda_\chi|^3)
\right],
\end{align}
which reproduces exactly the Drude contribution and the quadratic correction encoded in Eq.~(\ref{sigma_tensor_perturbative_main}). In particular, the coefficient of the $B^{2}$ term matches the combination of isotropic and anisotropic tensor structures, once the decomposition (\ref{J_FS_decomposition_main}) is taken into account.

This agreement provides a nontrivial consistency check of the formalism. It shows that the nonperturbative treatment correctly captures the standard weak-field physics while simultaneously extending it beyond the perturbative regime. At the same time, it clarifies the origin of apparent linear-in-$B$ contributions: these arise from the infrared sector of the continuum Weyl model and are absent in a strictly perturbative expansion, where the small-$k$ region is effectively regularized from the outset. Thus, the exact expressions derived here not only reproduce known results in the appropriate limit, but also reveal the limitations of the perturbative approach and the role of infrared physics in Weyl magnetotransport.

In this sense, the anisotropic conductivity provides a particularly clear diagnostic of nonperturbative semiclassical effects, as it isolates the infrared-sensitive sector of the theory and exposes its impact on observable transport coefficients.

{

\subsection{Finite-temperature, disorder, and lattice effects}

The infrared-sensitive behavior discussed above naturally raises the question of how additional physical scales may affect the transport response. In particular, finite temperature, momentum-dependent scattering processes, and microscopic lattice regularizations all introduce characteristic low-energy scales that could potentially compete with the infrared cutoff $k_{\rm IR}$. In this subsection we briefly discuss the robustness of our results against these effects.

The analysis presented above has been carried out at zero temperature, where the derivative of the Fermi-Dirac distribution reduces to a delta function at the Fermi surface. At finite temperature, the sharp Fermi-surface constraint is thermally broadened over an energy window of order $k_B T$. The transport coefficients may then be written in the generic form
\begin{equation}
\sigma_\chi (B,T)
=
\int_{-\infty}^{\infty} d\varepsilon\,
\Phi_\chi (\varepsilon,B)
\left(
-\frac{\partial f_0(\varepsilon)}{\partial \varepsilon}
\right),
\label{finite_T_kernel}
\end{equation}
where $\Phi_\chi (\varepsilon,B)$ is the regularized zero-temperature transport kernel evaluated at energy $\varepsilon$.

For $k_B T\ll \mu$, Eq.~\eqref{finite_T_kernel} admits a Sommerfeld expansion,
\begin{equation}
\sigma_\chi (B,T)
=
\Phi_\chi (\mu,B)
+
\frac{\pi^2}{6}(k_B T)^2
\left.
\frac{\partial^2 \Phi _\chi (\varepsilon,B)}
{\partial \varepsilon^2}
\right|_{\varepsilon=\mu}
+
\mathcal O(T^4),
\label{Sommerfeld_kernel}
\end{equation}
which can be equivalently written as
\begin{equation}
\sigma_\chi (B,T)
=
\sigma_\chi (B,0)
+
\frac{\pi^2}{6}(k_B T)^2
\left.
\frac{\partial^2 \Phi_\chi(\varepsilon,B)}
{\partial \varepsilon^2}
\right|_{\varepsilon=\mu}
+
\mathcal O(T^4).
\label{Sommerfeld_sigma}
\end{equation}

Equation~\eqref{Sommerfeld_sigma} shows that, in the low-temperature regime, finite temperature produces smooth corrections of order $T^2$ to the regularized zero-temperature result. Thermal broadening therefore softens the sharp Fermi-surface features quantitatively, but the infrared sensitivity discussed above originates from the singular low-energy behavior of the orbital magnetic moment, $m_{\alpha\mathbf{k}}\sim 1/k$, rather than from the zero-temperature form of the Fermi distribution itself. Consequently, within the regime $k_B T\ll \mu$, finite temperature does not provide a new infrared regularization mechanism and is not expected to qualitatively alter the nonanalytic magnetic-field trends discussed in this work. It is nevertheless useful to introduce the thermal momentum scale
\begin{equation}
k_T \sim \frac{k_B T}{\hbar v_F},
\label{thermal_scale}
\end{equation}
which characterizes the energy window over which thermal broadening becomes relevant. When the chemical potential lies very close to the Weyl node, $k_T$ may become comparable to other low-energy scales and contribute to the effective infrared cutoff. In such a situation, one may estimate the relevant low-energy scale as
\begin{equation}
k_{\rm eff}\sim \max(k_{\rm IR},k_T).
\label{effective_cutoff}
\end{equation}
However, throughout the regime considered in this work, where $k_B T\ll\mu$, one has $k_T\ll k_F$, the Sommerfeld expansion remains valid, and the dominant infrared regularization is expected to be controlled by $k_{\rm IR}$.

The robustness of the infrared behavior can also be examined by relaxing the constant-relaxation-time approximation. In realistic Weyl semimetals, different scattering mechanisms generally lead to momentum-dependent relaxation times. For example, short-range, Gaussian, and screened Coulomb impurities produce distinct energy dependences of $\tau(E)$, thereby modifying the relative contribution of low-energy quasiparticles to the transport coefficients. Such effects alter quantitative aspects of the response, including its overall magnitude and detailed scaling with the chemical potential, but they do not modify the underlying origin of the infrared sensitivity.

Indeed, the infrared behavior discussed in this work originates from the singular low-energy structure of the Weyl spectrum and, in particular, from the divergence of the orbital magnetic moment, $m_{\alpha\mathbf{k}}\sim 1/k$. We have verified that the standard impurity models considered above lead to different energy-dependent relaxation times \cite{PhysRevB.89.054202,PhysRevB.91.035202}, but none of them removes the infrared sensitivity of the corresponding transport integrals. Rather, within the semiclassical regime, disorder acts primarily as a momentum-dependent weighting factor through $\tau(E)$, while the dominant low-energy regularization remains controlled by $k_{\rm IR}$.

Disorder nevertheless introduces an additional energy scale associated with quasiparticle broadening,
\begin{equation}
\Gamma \sim \frac{\hbar}{\tau},
\end{equation}
which may become relevant as the system approaches the strong-disorder regime. In such a situation, one may expect an effective infrared scale determined by the competition between $k_{\rm IR}$, $k_T$, and the momentum scale associated with $\Gamma$. A quantitative analysis of this regime would require going beyond the weak-disorder semiclassical framework adopted here. Consequently, within the low-temperature and weak-disorder regime considered in this work, the nonanalytic magnetic-field dependence and the associated infrared behavior remain robust against both thermal broadening and realistic momentum-dependent scattering processes.

It is also instructive to consider the problem from the perspective of lattice regularizations of Weyl semimetals. A tight-binding description naturally regularizes the ultraviolet sector of the theory, but it does not eliminate the singular low-energy structure responsible for the infrared-sensitive response. For example, lattice Weyl models considered in Refs.~\cite{PhysRevB.93.035116,PhysRevB.103.115146} yield an orbital magnetic moment of the form
\begin{equation}
m_{\mathbf{k}}
\propto
\frac{\sin(ka)}{k^2},
\end{equation}
which, in the vicinity of the Weyl node ($ka\ll1$), reduces to
\begin{equation}
m_{\mathbf{k}}
\propto
\frac{a}{k}.
\end{equation}
Therefore, the lattice theory reproduces the same asymptotic low-energy behavior as the continuum Weyl Hamiltonian. This suggests that the infrared-sensitive transport response identified here should persist in microscopic lattice realizations, with the role of $k_{\rm IR}$ effectively played by the smallest physically accessible low-energy scale, such as finite-size effects, temperature, disorder broadening, or momentum resolution. From this perspective, lattice simulations provide a natural framework for investigating both the infrared sensitivity and the associated noncommutativity of limits discussed in the present work.

}

\section{Discussion} \label{conclusion}

In this work we have developed a nonperturbative semiclassical description of magnetotransport in Weyl semimetals, focusing on the Fermi-surface contribution to the conductivity in the presence of Berry curvature and orbital magnetic moment effects. By retaining the full magnetic-field dependence of the phase-space factor, generalized velocity, and band energy, we obtained closed-form expressions for both the isotropic and anisotropic components of the conductivity, valid to all orders in the magnetic field within the semiclassical regime.

A central outcome of our analysis is that the exact continuum formulation exhibits an intrinsic infrared sensitivity, originating from the singular behavior of the orbital magnetic moment, $\mathbf{m}_{\alpha\mathbf{k}}\sim 1/k$. This feature implies that the linearized Weyl Hamiltonian, when treated beyond leading order in the magnetic field, probes momentum regions where the semiclassical approximation ceases to be controlled. By enforcing the condition that orbital corrections remain small compared to the band energy, we identified a physically motivated infrared cutoff that regularizes the theory and ensures the internal consistency of the semiclassical approach.

Within this framework, we have shown that the weak-field expansion and the momentum integration do not commute in the continuum Weyl model. While perturbative treatments yield an analytic expansion starting at order $B^{2}$, the exact scalar transport coefficients contain nonanalytic contributions arising from the infrared sector. Although these terms cancel upon reconstructing the full conductivity tensor, they reveal that the continuum theory remains sensitive to low-momentum physics that is effectively excluded in conventional perturbative approaches.

This distinction leads naturally to the identification of two complementary transport regimes. In the conventional weak-field regime, $|\lambda_\chi|\ll1$, the magnetic field acts as a small perturbation and the transport response is well described by a quadratic magnetoconductivity. In contrast, when the dimensionless parameter $\lambda_\chi = B_\chi/B_{\mathrm{eff}}$ is not negligibly smal, either due to moderate magnetic fields or low carrier densities, the system enters a regime where magnetotransport becomes intrinsically nonperturbative. In this regime, the interplay between Berry curvature, orbital magnetic moment, and phase-space deformation leads to corrections that cannot be captured by a simple expansion in $B$.

Importantly, the anisotropic component of the conductivity provides a particularly clear diagnostic of these effects. Unlike the isotropic contribution, which remains infrared safe, the anisotropic response is directly sensitive to the low-momentum sector and therefore exhibits deviations from perturbative predictions even in the nominal weak-field limit. This makes it a promising observable for probing nonperturbative semiclassical effects in Weyl systems. {In this regard, our results complement recent studies emphasizing the role of the orbital magnetic moment in Berry-curvature-induced transport phenomena. While previous works have shown that orbital-moment corrections can strongly influence linear and nonlinear magnetoconductivities within weak-field approaches, the present analysis demonstrates that the orbital magnetic moment also controls the infrared structure of the transport problem when its full magnetic-field dependence is retained. This reveals a previously unexplored nonperturbative aspect of orbital-moment physics in Weyl semimetals.}

More broadly, our results clarify the limitations of perturbative semiclassical approaches and provide a unified framework that interpolates between weak-field transport and a regime dominated by nonperturbative orbital effects. The emergence of an infrared cutoff should be understood as a manifestation of the fact that any realistic Weyl material is ultimately defined on a lattice, which regularizes the low-energy sector. Within this perspective, the present theory captures the continuum limit while explicitly identifying the scales at which it breaks down. {We have also examined the effect of the Lorentz-force streaming term within the semiclassical Boltzmann equation. As discussed in Appendix~\ref{app:Lorentz}, this contribution generates the conventional transverse Hall response and modifies the conductivity perpendicular to the magnetic field through cyclotron-motion effects, while leaving the longitudinal magnetoconductivity and its infrared-sensitive nonperturbative structure unchanged. This demonstrates that the central conclusions of the present work originate from the geometric effects associated with Berry curvature and the orbital magnetic moment, rather than from conventional Lorentz-force corrections.}

Finally, the formalism developed here can be extended in several directions, including finite temperature, realistic band structures, and disorder beyond the relaxation-time approximation. Such generalizations would allow for a more direct connection with experiments and may help identify material platforms where the nonperturbative features discussed in this work can be observed.

\acknowledgements{ L.M.O. was supported by the SECIHTI PhD fellowship No. 834773. A.M.-R. acknowledges financial support by UNAM-PAPIIT project No. IG100224, UNAM-PAPIME project No. PE109226, by SECIHTI project No. CBF-2025-I-1862 and by the Marcos Moshinsky Foundation.}
 
\

\appendix

\section{Exact evaluation of the intrinsic current integral}
\label{appendix_intrinsic}

Let us consider the integral defined by Eq. (\ref{J_intrinsic_exact_2})
\begin{align}
I \equiv \int _{0} ^{\infty} dk \int _{-1} ^{1} d \xi \; \xi \; \Theta \left( \mu - \hbar v _{F} k - \frac{\chi e v _{F} B _{\chi} }{2k} \, \xi  \right).
\end{align}
Since $k>0$, it is convenient to perform the radial integral first. Defining
\begin{align}
k _{F} \equiv \frac{\mu}{\hbar v _{F}} > 0,
\qquad
b_\chi \equiv \frac{\chi e B_\chi}{2\hbar}, \label{definitions_appendix}
\end{align}
the argument of the step function can be rewritten as
\begin{align}
k ^{2} - k _{F} k + b _{\chi} \xi < 0 .
\end{align}
For fixed $\xi$, the corresponding roots are
\begin{align}
k _{\pm} ( \xi ) = \frac{k _{F} \pm \sqrt{k _{F} ^{2} - 4 b _{\chi} \xi}}{2}.
\end{align}
Hence, the radial integral is determined by the sign of $b _{\chi} \xi$. If $b _{\chi} \xi >0$, both roots are positive and the allowed interval is $k _{-} (\xi)<k<k _{+} (\xi)$, giving
\begin{align}
\int _{0} ^{\infty} dk\; \Theta \left( \mu - \hbar v _{F} k - \frac{\chi e v _{F} B _{\chi} }{2k} \xi \right) = k _{+} (\xi) - k _{-} (\xi) = \sqrt{ k _{F} ^{2} - 4 b _{\chi} \xi } .
\end{align}
By contrast, if $b _{\chi} \xi < 0$, only one root is positive and the allowed interval becomes $0<k<k_{+}(\xi)$, so that
\begin{align}
\int _{0} ^{\infty} dk\; \Theta \left( \mu - \hbar v _{F} k - \frac{\chi e v _{F} B _{\chi} }{2k} \xi \right) = k _{+} (\xi) = \frac{ k _{F} + \sqrt{k _{F} ^{2} - 4 b _{\chi} \xi}}{2} .
\end{align}
Therefore, the exact reduction is
\begin{align}
I &= \int _{-1} ^{1} d \xi \; \xi \; \Bigg[ \sqrt{k _{F} ^{2} - 4 b _{\chi} \xi } \, \Theta ( b _{\chi} \xi ) \, \Theta ( k _{F} ^{2} - 4 b _{\chi} \xi) + \frac{ k _{F} + \sqrt{k _{F} ^{2} - 4 b _{\chi} \xi}}{2} \, \Theta ( - b _{\chi} \xi) \, \Theta ( k _{F} ^{2} - 4 b _{\chi} \xi ) \Bigg] . \label{exact_reduction_appendix}
\end{align}
The first term in Eq.~\eqref{exact_reduction_appendix} is continuously connected to the usual Fermi-surface branch, whereas the second one comes from the complementary low-momentum sector of the linearized Weyl model. In order to obtain the full intrinsic current, both contributions must be retained. It is useful to analyze the two sectors separately.

We first consider the Fermi-surface contribution:
\begin{align}
I _{1} = \int _{-1} ^{1} d \xi \; \xi \, \sqrt{ k _{F} ^{2} - 4 b _{\chi} \xi } \; \Theta ( b _{\chi} \xi ) \; \Theta ( k _{F} ^{2} - 4 b _{\chi} \xi ) .
\end{align}
Introducing $\eta = \mathrm{sgn}( b_{\chi}) \, \xi$, this becomes
\begin{align}
I _{1} = \mathrm{sgn}(b _{\chi}) \int _{0} ^{\eta _{*}} d \eta \; \eta \, \sqrt{k _{F} ^{2} - 4 | b _{\chi}| \eta } , \qquad \eta _{*} = \min \left( 1 , \frac{k _{F} ^{2} }{ 4 |b _{\chi}| } \right) . \label{restriction}
\end{align}
Carrying out the integral, we obtain
\begin{align}
I _{1} = \mathrm{sgn}(b _{\chi} ) \; \frac{ k _{F} ^{5} - \bigl( k _{F} ^{2} + 6 \, |b _{\chi}| \, \eta _{*} \bigr) \bigl( k _{F} ^{2} - 4 \, |b _{\chi}| \, \eta _{*} \bigr) ^{3/2} }{ 60 \, |b _{\chi}| ^{2} } . 
\end{align}
From Eq. (\ref{definitions_appendix}) one finds $k_F^2/(4|b_\chi|)=\mu^2/(2e\hbar v_F^2 |B_\chi|)\gg 1$ in the semiclassical regime, where Landau quantization can be neglected and the wave-packet description remains valid, so that $\eta _{*} = 1$. The final result is then
\begin{align}
I _{1} = \mathrm{sgn}(b _{\chi} ) \; \frac{ k _{F} ^{5} - \bigl( k _{F} ^{2} + 6 \, |b _{\chi}|  \bigr) \bigl( k _{F} ^{2} - 4 \, |b _{\chi}| \bigr) ^{3/2} }{ 60 \, |b _{\chi}| ^{2} } .  \label{I1_result_appendix}
\end{align}
We now turn to the second branch, corresponding to $b_\chi \xi<0$. In this case it is convenient to write $\eta = - \, \mathrm{sgn}(b _{\chi} ) \, \xi$ so that again $\eta\in[0,1]$. Equation~\eqref{exact_reduction_appendix} then gives
\begin{align}
I _{2} = - \frac{1}{2} \, \mathrm{sgn}(b _{\chi} ) \int _{0} ^{1} d \eta \; \eta \, \left[ k _{F} + \sqrt{k _{F} ^{2} + 4 | b _{\chi} | \, \eta } \right] . \label{I2_eta_appendix}
\end{align}
Here no additional restriction analogous to Eq.~\eqref{restriction} is required, since the argument of the square root is always positive. The required integrals are quite simple. The result is
\begin{align}
I _{2} = - \mathrm{sgn} ( b _{\chi} )  \left[ \frac{ ( k _{F} ^{2} - 6 | b _{\chi} | ) ( k _{F} ^{2} + 4 | b _{\chi} | ) ^{3/2} - k _{F} ^{5} }{120 | b _{\chi} | ^{2} } + \frac{k _{F} }{4} \right] . \label{I2_result_appendix}
\end{align}
Adding Eqs.~\eqref{I1_result_appendix} and \eqref{I2_result_appendix}, we finally obtain the full result
\begin{align}
I = \mathrm{sgn}(b _{\chi} ) \left[ \frac{ k _{F} ^{5} -(k _{F} ^{2} + 6 | b _{\chi} | ) ( k _{F} ^{2} - 4 | b _{\chi} | ) ^{3/2} }{60 | b _{\chi} | ^{2} } + \frac{ k _{F} ^{5} - ( k _{F} ^{2} - 6 | b _{\chi} | )( k _{F} ^{2} + 4 | b _{\chi} | ) ^{3/2} }{120 | b _{\chi}| ^{2} } - \frac{k _{F}}{4} \right] . \label{I_exact_final_appendix}
\end{align}
Substituting this result in Eq.~(\ref{J_intrinsic_exact_2}) one gets the final result of Eq.~(\ref{J_intrinsic_exact_FIN}).

\section{Evaluation of the Fermi-surface contribution}
\label{app:FS}

In this appendix we present the derivation of the Fermi-surface contribution introduced in Eq.~(\ref{J_FS_2}). Since we focus on $\mu>0$, only the conduction band contributes, so that $s=+1$. The delta function imposes the constraint
\begin{align}
\hat{\mathbf{k}} \cdot \mathbf{B} _{\chi} = \frac{2 \chi k}{e v _{F} }\bigl( \mu - \hbar v _{F} k \bigr) ,
\end{align}
which enables a convenient rewriting of the generalized velocity, on the support of the integral, as
\begin{align}
\boldsymbol{\mathcal{V}} _{\alpha \mathbf{k}} = v _{F} \frac{ \hat{\mathbf{k}} }{\kappa}  (  3  \kappa -  2 ) +  v _{F} \, \Delta _{\chi} \, \frac{ \hat{\mathbf{B}} _{\chi}   }{ \kappa ^{3}} ( 1 -  \kappa ) , \label{Vrewritten_appendix}
\end{align}
where we introduced the dimensionless quantities $\kappa \equiv k / k _{F}$ and $ \Delta _{\chi} = \frac{\chi e B_\chi}{2\hbar k _{F} ^{2}}$. Therefore,
\begin{align}
    [ \boldsymbol{\mathcal{V}} _{\alpha \mathbf{k}} ] _{i} [ \boldsymbol{\mathcal{V}} _{\alpha \mathbf{k}} ] _{j} = v _{F} ^{2} \frac{ \hat{k} _{i} \hat{k} _{j} }{\kappa ^{2}}  (  3  \kappa -  2 ) ^{2} +  v _{F} ^{2} \, \Delta _{\chi} ^{2} \, \frac{ \hat{B} _{\chi i} \hat{B} _{\chi j}   }{ \kappa ^{6}} ( 1 -  \kappa ) ^{2} + v _{F} ^{2} \, \Delta _{\chi} (  3  \kappa -  2 )  ( 1 -  \kappa ) \, \frac{ \hat{k} _{i} \hat{B} _{\chi j} + \hat{k} _{j} \hat{B} _{\chi i}  }{ \kappa ^{4} } .
\end{align}
In a similar fashion, the Berry-curvature correction to the phase-space measure becomes
\begin{align}
D _{\alpha \mathbf{k}} = \frac{\kappa}{2 \kappa - 1} . \label{Dkappa_appendix}
\end{align}
Substituting Eqs.~(\ref{Vrewritten_appendix}) and (\ref{Dkappa_appendix}) into Eq.~(\ref{J_FS_2}), and using
\begin{align}
\delta \left( \mu - \hbar v _{F} k - \frac{\chi e v _{F}}{2k} \, \hat{\mathbf{k}} \cdot \mathbf{B} _{\chi}  \right)
=
\frac{\kappa}{\hbar v_F k_F}\,
\delta \left(  \kappa ( 1 -  \kappa ) - \Delta _{\chi} \, \hat{\mathbf{k}} \cdot \hat{\mathbf{B}} _{\chi}  \right) ,
\end{align}
we obtain
\begin{align}
J _{i} ^{\mathrm{FS}} &= \frac{ e ^{2} v _{F}  \tau }{4 \pi ^{2} h k _{F} } \sum _{\chi } E _{\chi j} \int  \frac{d ^{3} \mathbf{k}}{2 \kappa - 1} \Bigg\{ (  3  \kappa -  2 ) ^{2} \, \hat{k} _{i} \hat{k} _{j} + \Delta _{\chi} ^{2} \, \frac{ \hat{B} _{\chi i} \hat{B} _{\chi j} }{ \kappa ^{4}} \, ( 1 -  \kappa ) ^{2} \notag\\ &\hspace{5.3cm} + \Delta _{\chi} \, (  3  \kappa -  2 )  ( 1 -  \kappa ) \, \frac{ \hat{k} _{i} \hat{B} _{\chi j} + \hat{k} _{j} \hat{B} _{\chi i}  }{ \kappa ^{2} } \Bigg\} \; \delta \left(  \kappa ( 1 -  \kappa ) - \Delta _{\chi} \, \hat{\mathbf{k}} \cdot \hat{\mathbf{B}} _{\chi}  \right) . \label{JFS_kappa_appendix}
\end{align}
The evaluation of Eq.~(\ref{JFS_kappa_appendix}) reduces to the following dimensionless angular integrals:
\begin{align}
    Q(\kappa ) &= \int d \Omega \;  \delta \left(  \kappa ( 1 -  \kappa ) - \Delta _{\chi} \, \hat{\mathbf{k}} \cdot \hat{\mathbf{B}} _{\chi}  \right) , \label{Q_escalar} \\
    Q _{i}(\kappa ) &= \int d \Omega \; \hat{k} _{i} \; \delta \left(  \kappa ( 1 -  \kappa ) - \Delta _{\chi} \, \hat{\mathbf{k}} \cdot \hat{\mathbf{B}} _{\chi}  \right) , \label{Qi_vector} \\
    Q _{ij}(\kappa ) &= \int d \Omega \; \hat{k} _{i} \hat{k} _{j} \; \delta \left(  \kappa ( 1 -  \kappa ) - \Delta _{\chi} \, \hat{\mathbf{k}} \cdot \hat{\mathbf{B}} _{\chi}  \right) . \label{Qij_tensor}
\end{align}
These are evaluated in the following. To simplify the angular integrals, we align the polar axis with $\hat{\mathbf{B}} _{\chi}$ and define
\begin{align}
\xi \equiv \hat{\mathbf{k}} \cdot \hat{\mathbf{B}} _{\chi} = \cos \theta ,
\end{align}
so that $d \Omega = 2 \pi \, d \xi$. In this representation, the argument of the delta function depends only on $\xi$, enabling a direct evaluation of the integrals. Then, the scalar fucntion (\ref{Q_escalar}) becomes
\begin{align}
    Q(\kappa ) &= 2 \pi \int _{-1} ^{1} d \xi \;  \delta \left(  \kappa ( 1 -  \kappa ) - \Delta _{\chi} \, \xi  \right) = \frac{2 \pi}{ | \Delta _{\chi} | } \Theta \left( | \Delta _{\chi} | - | \kappa ( 1 - \kappa ) | \right) .    \label{Q_escalar_2}
\end{align}
By rotational symmetry around the axis defined by $\hat{\mathbf B}_\chi$, the only vector that can result from the angular integral is $\hat{\mathbf{B}} _{\chi}$ itself. Therefore, the $Q _{i} ( \kappa ) $ defined by Eq. (\ref{Qi_vector}) must be proportional to $\hat{B} _{\chi i}$, and we write
\begin{align}
Q _{i} ( \kappa ) = A _{\chi} ( \kappa ) \, \hat{B} _{\chi i} .
\end{align}
To determine the scalar coefficient $A _{\chi} ( \kappa ) $, we contract with $\hat{B} _{\chi i}$:
\begin{align}
    A _{\chi} ( \kappa ) &= \int d \Omega \; \hat{\mathbf{k}} \cdot \hat{\mathbf{B}} _{\chi} \; \delta \left(  \kappa ( 1 -  \kappa ) - \Delta _{\chi} \, \hat{\mathbf{k}} \cdot \hat{\mathbf{B}} _{\chi}  \right) . 
\end{align}
This is now a scalar integral, which can be easily evaluated using the same coordinate system as before. We obtain
\begin{align}
    A _{\chi} ( \kappa ) &= 2 \pi \int _{-1} ^{1} d \xi \; \xi \; \delta \left(  \kappa ( 1 -  \kappa ) - \Delta _{\chi} \, \xi \right) = 2 \pi \, \mathrm{sgn} ( \Delta _{\chi} ) \, \frac{ \kappa ( 1 - \kappa ) }{ | \Delta _{\chi} | ^{2} } \Theta \left( | \Delta _{\chi} | - | \kappa ( 1 - \kappa ) | \right) . \label{Qi_kappa_final}
\end{align}
We now evaluate the tensor form $Q_{ij}$ defined by Eq. (\ref{Qij_tensor}). Owing to the rotational symmetry around $\hat{\mathbf{B}} _{\chi}$, this tensor can be expressed as
\begin{align}
Q _{ij} ( \kappa ) = C _{\chi} ( \kappa ) \; \delta _{ij} + D _{\chi} ( \kappa ) \; \hat{B}_{\chi i} \hat{B}_{\chi j} . \label{Qij_decomposition_appendix}
\end{align}
Taking the trace of Eq.~(\ref{Qij_decomposition_appendix}), we obtain
\begin{align}
Q ( \kappa ) = Q _{ii} ( \kappa ) = 3 C _{\chi}( \kappa ) + D _{\chi} ( \kappa ) , \label{trace_Qij_appendix}
\end{align}
where $Q ( \kappa )$ is the previously evaluated scalar function. On the other hand, contracting Eq.~(\ref{Qij_decomposition_appendix}) with
$\hat B_{\chi i}\hat B_{\chi j}$ gives
\begin{align}
Q_{\parallel}(\kappa)
\equiv
\hat B_{\chi i}\hat B_{\chi j}Q_{ij}(\kappa)
=
C_{\chi}(\kappa)+D_{\chi}(\kappa),
\label{longitudinal_Qij_appendix}
\end{align}
with
\begin{align}
Q _{\parallel} ( \kappa ) = 2 \pi  \int _{-1} ^{1} d \xi \; \xi ^{2} \, \delta \left( \kappa ( 1 -\kappa ) - \Delta _{\chi} \, \xi \right) . \label{Qparallel_appendix}
\end{align}
Solving Eqs.~(\ref{trace_Qij_appendix}) and (\ref{longitudinal_Qij_appendix}), we find
\begin{align}
C_{\chi}(\kappa)
&=
\frac{1}{2}\Big[\,Q(\kappa)-Q_{\parallel}(\kappa)\,\Big],
\label{Cchi_integral_appendix}
\\
D_{\chi}(\kappa)
&=
\frac{1}{2}\Big[\,3Q_{\parallel}(\kappa)-Q(\kappa)\,\Big].
\label{Dchi_integral_appendix}
\end{align}
Evaluating the integral in the same manner, we find
\begin{align}
Q _{\parallel} ( \kappa )
&= 2 \pi  \int _{-1} ^{1} d \xi \; \xi ^{2} \, \delta \left( \kappa ( 1 -\kappa ) - \Delta _{\chi} \, \xi \right) = 2\pi\,
\frac{\kappa^2(1-\kappa)^2}{|\Delta_\chi|^3}
\Theta \left(|\Delta_\chi|-|\kappa(1-\kappa)|\right).
\end{align}
Using Eqs.~(\ref{Q_escalar_2}) and (\ref{Qparallel_appendix}), the scalar functions
$C_\chi(\kappa)$ and $D_\chi(\kappa)$ follow directly from
Eqs.~(\ref{Cchi_integral_appendix}) and (\ref{Dchi_integral_appendix}). Substituing these expressions back into Eq. (\ref{JFS_kappa_appendix}), the Fermi-surface current can be decomposed as
\begin{align}
\mathbf{J} ^{\mathrm{FS}} = \sum _{\chi} \left[ \sigma _{\chi} (B) \, \mathbf{E} _{\chi} + \bar{\sigma} _{\chi} (B) \, ( \mathbf{E} _{\chi} \cdot \hat{\mathbf{B}} _{\chi} ) \, \hat{\mathbf{B}} _{\chi} \right] , \label{J_FS_decomposition_app}
\end{align}
where
\begin{align}
    \sigma _{\chi} (B) = \frac{ e ^{2} v _{F}  \tau k _{F} ^{2}}{4 \pi ^{2} h  } \int _{0} ^{\infty} d \kappa \;  \frac{\kappa ^{2} \, (  3  \kappa -  2 ) ^{2}  }{2 \kappa - 1}    \, C _{\chi} ( \kappa )      
\end{align}
and
\begin{align}
\bar{\sigma} _{\chi} (B) &= \frac{ e ^{2} v _{F}  \tau k _{F} ^{2}}{4 \pi ^{2} h  } \int _{0} ^{\infty}  \frac{\kappa ^{2} \, d \kappa }{2 \kappa - 1} \Bigg\{  (  3  \kappa -  2 ) ^{2} \, D _{\chi} ( \kappa )   + \Delta _{\chi} ^{2} \, \frac{ ( 1 -  \kappa ) ^{2} }{ \kappa ^{4}} \,  Q (\kappa ) + \Delta _{\chi} \,  \frac{ 2 (  3  \kappa -  2 )  ( 1 -  \kappa )   }{ \kappa ^{2} } A _{\chi} ( \kappa ) \Bigg\}   . \label{JFS_kappa_appendix_2}
\end{align}
Substituting the coefficients $C _{\chi}$, $D _{\chi} $, $Q$ and $A _{\chi}$ we establishes eqs. (\ref{sigma_iso_main}) and (\ref{sigmaB_exact}) of the main text.

{

\section{Lorentz-force streaming term}
\label{app:Lorentz}

In this Appendix we discuss how the Fermi-surface contribution is modified when the Lorentz-force streaming term is retained in the Boltzmann equation. This term is not included explicitly in Eq.~(\ref{dist_function}) of the main text, where the magnetic-field dependence enters through the phase-space factor, the generalized velocity, and the orbital-magnetic-moment correction to the band energy. The analysis below shows that the Lorentz-force streaming term generates a conventional transverse Hall contribution and renormalizes the transverse conductivity, while leaving the strictly longitudinal response unchanged.

For a homogeneous stationary system, we write
\begin{equation}
f_{\alpha\mathbf{k}}
=
f_0(\mathcal{E}_{\alpha\mathbf{k}})
+
\delta f_{\alpha\mathbf{k}},
\end{equation}
where \(\delta f_{\alpha\mathbf{k}}\) is linear in the electric field. Keeping the magnetic part of the semiclassical acceleration acting on \(\delta f_{\alpha\mathbf{k}}\), the relaxation-time Boltzmann equation becomes
\begin{equation}
\delta f_{\alpha\mathbf{k}}
+
\tau
\dot{\mathbf{k}}_{\alpha}
\cdot
\nabla_{\mathbf{k}}
\delta f_{\alpha\mathbf{k}}
=
e\tau
D_{\alpha\mathbf{k}}
\left(
\mathbf{E}_{\chi}\cdot
\mathbf{V}_{\alpha\mathbf{k}}
\right)
\frac{\partial f_0}
{\partial \mathcal{E}_{\alpha\mathbf{k}}},
\label{eq:Lorentz_Boltzmann}
\end{equation}
where
\begin{equation}
\dot{\mathbf{k}}_{\alpha}
=
-\frac{e}{\hbar}
D_{\alpha\mathbf{k}}\,
\mathbf{v}_{\alpha\mathbf{k}}
\times
\mathbf{B}_{\chi}.
\label{eq:Lorentz_kdotB}
\end{equation}
We choose the magnetic field along the $z$ direction, $\mathbf{B}_{\chi}
=
B_{\chi}\hat{\mathbf{z}}$, and introduce spherical coordinates, $\mathbf{k}
=
k
\left(
\sin\theta\cos\phi,
\sin\theta\sin\phi,
\cos\theta
\right)$. For an axially symmetric dispersion around \(\mathbf{B}_{\chi}\), the band velocity has no azimuthal component and can be written as
\begin{equation}
\mathbf{v}_{\alpha\mathbf{k}}
=
v_{\alpha r}(k,\theta)\hat{\mathbf{k}}
+
v_{\alpha\theta}(k,\theta)\hat{\boldsymbol{\theta}} .
\end{equation}
such that
\begin{equation}
\dot{\mathbf{k}}_{\alpha}
\cdot
\nabla_{\mathbf{k}}
=
\omega_{\alpha}(k,\theta)
\frac{\partial}{\partial\phi},
\label{eq:Lorentz_streaming}
\end{equation}
with
\begin{equation}
\omega_{\alpha}(k,\theta)
=
\frac{eB_{\chi}}{\hbar}
D_{\alpha\mathbf{k}}
\frac{
v_{\alpha r}(k,\theta)\sin\theta
+
v_{\alpha\theta}(k,\theta)\cos\theta
}
{k\sin\theta}.
\label{eq:Lorentz_omega_general}
\end{equation}
Thus the Lorentz-force term acts as an azimuthal streaming operator.

We now decompose the generalized velocity (\ref{generalized_velocity}) as
\begin{equation}
\boldsymbol{\mathcal V} _{\alpha\mathbf{k}}
=
{\mathcal V} _{\alpha r}(k,\theta)\hat{\mathbf{k}}
+
{\mathcal V} _{\alpha B}(k,\theta)\hat{\mathbf{B}}_{\chi}.
\label{eq:Lorentz_Vdecomp}
\end{equation}
Then
\begin{align}
\mathbf{E}_{\chi}\cdot \boldsymbol{\mathcal V} _{\alpha\mathbf{k}}
=
&
E_{\chi,z}
\left[
{\mathcal V} _{\alpha r}(k,\theta)\cos\theta
+
{\mathcal V} _{\alpha B}(k,\theta)
\right] +
{\mathcal V} _{\alpha r}(k,\theta)\sin\theta
\left(
E_{\chi,x}\cos\phi
+
E_{\chi,y}\sin\phi
\right).
\label{eq:Lorentz_EdotV}
\end{align}
The longitudinal part is independent of \(\phi\), whereas the transverse part contains only the \(m=\pm 1\) azimuthal harmonics. Defining
\begin{equation}
\gamma_{\alpha}(k,\theta)
=
\tau\omega_{\alpha}(k,\theta),
\end{equation}
the solution of Eq.~\eqref{eq:Lorentz_Boltzmann} is
\begin{align}
\delta f_{\alpha\mathbf{k}}
=
&
e\tau
D_{\alpha\mathbf{k}} \,  
\frac{\partial f_0}
{\partial \mathcal{E}_{\alpha\mathbf{k}}} \, 
\Bigg\{
E_{\chi,z}
\left[
{\mathcal V} _{\alpha r}\cos\theta
+
{\mathcal V} _{\alpha B}
\right]
+
{\mathcal V} _{\alpha r}\sin\theta
\frac{
\left(E_{\chi,x}-\gamma_{\alpha}E_{\chi,y}\right)\cos\phi
+
\left(E_{\chi,y}+\gamma_{\alpha}E_{\chi,x}\right)\sin\phi
}
{1+\gamma_{\alpha}^{2}}
\Bigg\}.
\label{eq:Lorentz_df}
\end{align}
All functions \({\mathcal V}_{\alpha r}\), \({\mathcal V}_{\alpha B}\), and \(\gamma_{\alpha}\) are evaluated at \((k,\theta)\). The Fermi-surface current is
\begin{equation}
\mathbf{J}^{\rm FS}
=
-e
\sum_{\alpha}
\int
\frac{d^3 \mathbf{k} }{(2\pi)^3}
\boldsymbol{\mathcal V} _{\alpha\mathbf{k}}\,
\delta f_{\alpha\mathbf{k}} .
\label{eq:Lorentz_Jstart}
\end{equation}
At zero temperature,
\begin{equation}
\frac{\partial f_0}
{\partial \mathcal{E}_{\alpha\mathbf{k}}}
=
-\delta
\left(
\mu-\mathcal{E}_{\alpha\mathbf{k}}
\right).
\end{equation}
Substitution of Eq.~\eqref{eq:Lorentz_df} into Eq.~\eqref{eq:Lorentz_Jstart}, followed by the integration over \(\phi\), yields
\begin{equation}
\mathbf{J}^{\rm FS}
=
\sigma_{\perp}^{L}(B)
\mathbf{E}_{\chi,\perp}
+
\sigma_{\parallel}^{L}(B)
\mathbf{E}_{\chi,\parallel}
+
\sigma_{H}^{L}(B)
\hat{\mathbf{B}}_{\chi}
\times
\mathbf{E}_{\chi,\perp},
\label{eq:Lorentz_tensor}
\end{equation}
where
\begin{equation}
\mathbf{E}_{\chi,\parallel}
= ( \mathbf{E}_{\chi}\cdot\hat{\mathbf{B}}_{\chi} )
\hat{\mathbf{B}}_{\chi},
\qquad
\mathbf{E}_{\chi,\perp}
=
\mathbf{E}_{\chi}
-
\mathbf{E}_{\chi,\parallel}.
\end{equation}
The corresponding conductivities are
\begin{equation}
\sigma_{\parallel}^{L}(B)
=
e^2\tau
\sum_{\alpha}
\int
\frac{d^3 \mathbf{k} }{(2\pi)^3}
D_{\alpha\mathbf{k}}
\left[
{\mathcal V} _{\alpha r}\cos\theta
+
{\mathcal V} _{\alpha B}
\right]^2
\delta
\left(
\mu-\mathcal{E}_{\alpha\mathbf{k}}
\right),
\label{eq:Lorentz_sigmapar}
\end{equation}
\begin{equation}
\sigma_{\perp}^{L}(B)
=
\frac{e^2\tau}{2}
\sum_{\alpha}
\int
\frac{d^3 \mathbf{k} }{(2\pi)^3}
D_{\alpha\mathbf{k}}
{\mathcal V} _{\alpha r}^{2}
\sin^2\theta
\frac{1}{1+\gamma_{\alpha}^{2}}
\delta
\left(
\mu-\mathcal{E}_{\alpha\mathbf{k}}
\right),
\label{eq:Lorentz_sigmaperp}
\end{equation}
and
\begin{equation}
\sigma_{H}^{L}(B)
=
\frac{e^2\tau}{2}
\sum_{\alpha}
\int
\frac{d^3 \mathbf{k} }{(2\pi)^3}
D_{\alpha\mathbf{k}}
{\mathcal V} _{\alpha r}^{2}
\sin^2\theta
\frac{\gamma_{\alpha}}{1+\gamma_{\alpha}^{2}}
\delta
\left(
\mu-\mathcal{E}_{\alpha\mathbf{k}}
\right).
\label{eq:Lorentz_sigmaH}
\end{equation}
Therefore, the Lorentz-force streaming term leaves the longitudinal conductivity unchanged, renormalizes the transverse conductivity through the factor \(1/(1+\gamma_{\alpha}^{2})\), and generates an additional Hall contribution proportional to \(\gamma_{\alpha}/(1+\gamma_{\alpha}^{2})\).

We now specialize the above expressions to the Weyl model considered in the main text. We restrict to the conduction band and use the dimensionless variables
\begin{equation}
\kappa=\frac{k}{k_F},
\qquad
\Delta_{\chi}
=
\frac{\chi eB_{\chi}}{2\hbar k_F^2}.
\end{equation}
On the support of the energy-conservation constraint, the generalized velocity can be written as
\begin{equation}
\boldsymbol{\mathcal V} _{\alpha\mathbf{k}}
=
v_F
\frac{3\kappa-2}{\kappa}
\hat{\mathbf{k}}
+
v_F
\frac{\Delta_{\chi}(1-\kappa)}{\kappa^3}
\hat{\mathbf{B}}_{\chi}.
\label{eq:Lorentz_WeylV}
\end{equation}
Furthermore,
\begin{equation}
D_{\alpha\mathbf{k}}
=
\frac{\kappa}{2\kappa-1}.
\label{eq:Lorentz_WeylD}
\end{equation}
For the field-corrected Weyl dispersion, the band-velocity components entering Eq.~\eqref{eq:Lorentz_omega_general} give, on the same constraint surface,
\begin{equation}
v_{\alpha r}\sin\theta+v_{\alpha\theta}\cos\theta
=
v_F\sin\theta
\frac{3\kappa-2}{\kappa}.
\end{equation}
Therefore,
\begin{equation}
\omega_{\chi}(\kappa)
=
\frac{eB_{\chi}v_F}{\hbar k_F}
\frac{3\kappa-2}{\kappa(2\kappa-1)},
\label{eq:Lorentz_WeylOmega}
\end{equation}
and
\begin{equation}
\gamma_{\chi}(\kappa)
=
\tau
\frac{eB_{\chi}v_F}{\hbar k_F}
\frac{3\kappa-2}{\kappa(2\kappa-1)}.
\label{eq:Lorentz_WeylGamma}
\end{equation}

Using the angular integrals of Appendix B, the transverse conductivity including the Lorentz-force streaming term becomes
\begin{align}
\sigma_{\perp,\chi}^{L}(B)
=
&
\frac{e^2v_F\tau k_F^2}{4\pi h}
\int_0^\infty
d\kappa\,
\frac{\kappa^2(3\kappa-2)^2}{2\kappa-1}
\Theta
\left(
|\lambda_{\chi}|
-
\kappa|1-\kappa|
\right)
\left[
\frac{1}{|\lambda_{\chi}|}
-
\frac{\kappa^2(1-\kappa)^2}{|\lambda_{\chi}|^3}
\right]
\frac{1}{1+\gamma_{\chi}^{2}(\kappa)} .
\label{eq:Lorentz_WeylSigmaperp}
\end{align}
Similarly, the Hall conductivity generated by the Lorentz-force streaming term is
\begin{align}
\sigma_{H,\chi}^{L}(B)
=
&
\frac{e^2v_F\tau k_F^2}{4\pi h}
\int_0^\infty
d\kappa\,
\frac{\kappa^2(3\kappa-2)^2}{2\kappa-1}
\Theta
\left(
|\lambda_{\chi}|
-
\kappa|1-\kappa|
\right) 
\left[
\frac{1}{|\lambda_{\chi}|}
-
\frac{\kappa^2(1-\kappa)^2}{|\lambda_{\chi}|^3}
\right]
\frac{\gamma_{\chi}(\kappa)}{1+\gamma_{\chi}^{2}(\kappa)} .
\label{eq:Lorentz_WeylSigmaH}
\end{align}
The longitudinal conductivity is not modified by the Lorentz-force streaming term and is given by
\begin{equation}
\sigma_{\parallel,\chi}^{L}(B)
=
\sigma_{\chi}(B)
+
\bar{\sigma}_{\chi}(B),
\label{eq:Lorentz_WeylSigmapar}
\end{equation}
where \(\sigma_{\chi}(B)\) and \(\bar{\sigma}_{\chi}(B)\) are the coefficients derived in the main text. Equations~\eqref{eq:Lorentz_WeylSigmaperp}-\eqref{eq:Lorentz_WeylSigmapar} show explicitly how the result of the main text is recovered in the weak-cyclotron limit,
\begin{equation}
|\gamma_{\chi}(\kappa)|\ll1.
\end{equation}
In this limit,
\begin{equation}
\sigma_{\perp,\chi}^{L}(B)
\to
\sigma_{\chi}(B),
\qquad
\sigma_{H,\chi}^{L}(B)
\to
0,
\qquad
\sigma_{\parallel,\chi}^{L}(B)
\to
\sigma_{\chi}(B)+\bar{\sigma}_{\chi}(B).
\end{equation}
Moreover, for the strictly longitudinal geometry,
\begin{equation}
\mathbf{E}_{\chi}\parallel \mathbf{B}_{\chi},
\end{equation}
one has \(\mathbf{E}_{\chi,\perp}=0\), and the transverse and Hall terms in Eq.~\eqref{eq:Lorentz_tensor} do not contribute. Hence, the longitudinal magnetoconductivity analyzed in the main text is unaffected by the explicit Lorentz-force streaming term.

}

\section{Perturbative weak-field expansion of the Fermi-surface conductivity tensor}
\label{app:perturbative_tensor}

For completeness, in this Appendix we reproduce the standard weak-field expansion of the Fermi-surface conductivity tensor within the semiclassical framework. Although this calculation is implicit in previous perturbative treatments of Weyl magnetotransport, it is useful to present it here in our notation in order to facilitate the comparison with the nonperturbative expressions derived in the main text.

We start from the Fermi-surface current in Eq.~(\ref{J_FS}), $J ^{\mathrm{FS}} _{i} = \sum _{\chi} \sigma ^{(\chi)} _{ij} E _{\chi,j}$, with
\begin{align}
\sigma ^{(\chi)} _{ij} = - e ^{2} \tau \int \frac{d ^{3}\mathbf{k}}{( 2 \pi ) ^{3}} \; D _{\chi \mathbf{k}} \, \mathcal{V} _{\chi \mathbf{k},i} \, \mathcal{V} _{\chi \mathbf{k},j} \, \frac{\partial f _{0}(\mathcal{E} _{\chi \mathbf{k}})}{\partial \mathcal{E} _{\chi \mathbf{k}}} . \label{app:sigma_tensor_start}
\end{align}
Here, all field dependence is contained in the phase-space factor $D_{\chi\mathbf{k}}$, the generalized velocity $\boldsymbol{\mathcal V}_{\chi\mathbf{k}}$, and the field-corrected band energy $\mathcal E_{\chi\mathbf{k}}$.

We restrict to the conduction band, $\mu>0$, for which the zero-field energy and velocity are $E _{\chi\mathbf{k}}=\hbar v_{F}k $ and $
v _{i}=v_{F}\hat{k}_{i}$, respectively. For the linearized Weyl Hamiltonian, the Berry curvature and orbital magnetic moment are $\Omega_{\chi,i} = - \chi\, \hat{k}_{i} / 2k^{2}$ and  $m_{\chi,i} = -\chi\, (e v_{F} / 2k ) \,\hat{k}_{i}$. Accordingly, the phase-space factor admits the weak-field expansion (up to quadratic orde in the magnetic field)
\begin{align}
D _{\chi\mathbf{k}} & \approx 1 + \chi\,\frac{e}{2\hbar}\, \frac{\hat{\mathbf k}\cdot\mathbf B_{\chi}}{k^{2}} + \frac{e^{2}}{4\hbar^{2}}\, \frac{(\hat{\mathbf k}\cdot\mathbf B_{\chi})^{2}}{k^{4}} + \mathcal O(B_{\chi}^{3}). \label{app:D_expand}
\end{align}
Next, we consider the generalized velocity. For the Weyl model, the exact expression is given by Eq. (\ref{generalized_velocity_exact_final}). Hence, in the perturbative expansion one must write $\mathcal V_i = \mathcal V_i^{(0)} + \mathcal V_i^{(1)} + \mathcal V_i^{(2)}$
with
\begin{align}
\mathcal V_i^{(0)} &= v_F \hat{k}_i, \qquad \mathcal V_i^{(1)} = -\chi\,\frac{e v_F}{\hbar}\, \frac{(\hat{\mathbf k}\cdot\mathbf B_{\chi})\hat{k}_i}{k^2}, \qquad \mathcal V_i^{(2)} = \frac{e^2 v_F}{4\hbar^2}\, \frac{(\hat{\mathbf k}\cdot\mathbf B_{\chi}) B_{\chi i}}{k^4}.
\end{align}
Expanding the product $\mathcal{V} _{i} \mathcal{V}_{j}$ up to second order in the magnetic field, we obtain
\begin{align}
\mathcal{V} _{i} \mathcal{V} _{j} &= v _{F} ^{2} \, \hat{k} _{i} \hat{k} _{j} - 2 \chi \, \frac{e v _{F} ^{2}}{\hbar} \, \frac{(\hat{\mathbf{k}} \cdot \mathbf{B} _{\chi})}{k ^{2}} \, \hat{k} _{i} \hat{k} _{j}  + \frac{e ^{2} v _{F} ^{2}}{\hbar ^{2}} \, \frac{(\hat{\mathbf{k}} \cdot \mathbf{B}_{\chi}) ^{2}}{k ^{4}} \, \hat{k} _{i} \hat{k} _{j} + \frac{e ^{2} v _{F} ^{2}}{4 \hbar ^{2}} \, \frac{(\hat{\mathbf{k}} \cdot \mathbf{B} _{\chi})}{k ^{4}} \, \left( \hat{k} _{i} B _{\chi j} + B _{\chi i} \hat{k} _{j} \right) + \mathcal{O} (B _{\chi} ^{3} ) . \label{app:VV_expand}
\end{align}
Finally, we expand the derivative of the Fermi distribution around the zero-field energy:
\begin{align}
\frac{\partial f_{0}(\mathcal E_{\chi\mathbf{k}})}{\partial \mathcal E_{\chi\mathbf{k}}}
=
f_{0}'(E_{\chi\mathbf{k}})
-
(\mathbf m_{\chi\mathbf{k}}\cdot\mathbf B_{\chi})\,f_{0}''(E_{\chi\mathbf{k}})
+
\frac12(\mathbf m_{\chi\mathbf{k}}\cdot\mathbf B_{\chi})^{2}
f_{0}'''(E_{\chi\mathbf{k}})
+
\mathcal O(B_{\chi}^{3}).
\label{app:f_expand}
\end{align}
At zero temperature,
\begin{align}
f_{0}'(\varepsilon)=-\delta(\mu-\varepsilon),
\qquad
f_{0}''(\varepsilon)=\delta'(\mu-\varepsilon),
\qquad
f_{0}'''(\varepsilon)=-\delta''(\mu-\varepsilon).
\label{app:delta_derivatives}
\end{align}
Substituting Eqs.~(\ref{app:D_expand}), (\ref{app:VV_expand}), and (\ref{app:f_expand}) into Eq.~(\ref{app:sigma_tensor_start}), and collecting terms order by order, the conductivity tensor can be written as
\begin{align}
\sigma^{(\chi)}_{ij} = \sigma^{(\chi,0)}_{ij} + \sigma^{(\chi,1)}_{ij} + \sigma^{(\chi,2)}_{ij} + \mathcal{O} (B _{\chi} ^{3}). \label{total_conductivity}
\end{align}
The zero-field contribution is the usual Drude term,
\begin{align}
\sigma^{(\chi,0)}_{ij}
&=
e^{2}\tau
\int \frac{d^{3}\mathbf{k}}{(2\pi)^{3}}\;
v_F^2 \hat{k}_i\hat{k}_j\,
\delta(\mu-\hbar v_F k) =
\frac{e^{2}v_F\tau k_F^{2}}{3\pi h}\,\delta_{ij},
\label{app:zero_field_tensor}
\end{align}
where $k_F=\frac{\mu}{\hbar v_F}$. At first order in the magnetic field, all contributions vanish after angular integration. Indeed, every linear term is proportional to an odd power of \(\hat{\mathbf k}\cdot\mathbf B_{\chi}\), and therefore integrates to zero over the Fermi sphere. Hence,
\begin{align}
\sigma^{(\chi,1)}_{ij}=0.
\label{app:linear_vanish}
\end{align}
At second order, the surviving terms originate from three distinct sources: the quadratic term in the phase-space factor, the mixed products in \(\mathcal V_i\mathcal V_j\), and the second-order expansion of the equilibrium distribution. After straightforward algebra, the result can be organized in the rotationally covariant form
\begin{align}
\sigma ^{(\chi,2)} _{ij} = A _{\chi} \, B _{\chi} ^{2} \, \delta_{ij} + C _{\chi} \, B _{\chi i} B _{\chi j} . \label{app:second_order_structure}
\end{align}
The required angular averages are
\begin{align}
\int d\Omega\;\hat{k}_{i}\hat{k}_{j}
&=
\frac{4\pi}{3}\,\delta_{ij}, \qquad \quad
\int d\Omega\;\hat{k}_{i}\hat{k}_{j}\hat{k}_{l}\hat{k}_{m}
=
\frac{4\pi}{15}
\left(
\delta_{ij}\delta_{lm}
+
\delta_{il}\delta_{jm}
+
\delta_{im}\delta_{jl}
\right).
\label{app:angular_averages}
\end{align}
Performing the radial integrals generated by \(\delta(\mu-\hbar v_F k)\), \(\delta'(\mu-\hbar v_F k)\), and \(\delta''(\mu-\hbar v_F k)\), one finds
\begin{align}
A_{\chi}
=
-
\frac{e^{4}v_F\tau}{30\pi h\,\hbar^{2}k_F^{2}},
\qquad
C_{\chi}
=
\frac{e^{4}v_F\tau}{10\pi h\,\hbar^{2}k_F^{2}}.
\end{align}
Substituting these coefficients into Eq. (\ref{app:second_order_structure}) the total conductivity tensor (\ref{total_conductivity}) per chirality can be written as
\begin{align}
\sigma_{ij}^{(\chi)}
=
\frac{e^{2}v_F\tau k_F^{2}}{3\pi h}
\left[
\left(1-\frac{2}{5}\lambda_{\chi}^{2}\right)\delta_{ij}
+
\frac{6}{5}\lambda_{\chi}^{2}\,
\hat B_{\chi i}\hat B_{\chi j}
\right]
+
\mathcal O(\lambda_{\chi}^{4}),
\label{app:perturbative_final_tensor}
\end{align}
where we used the definitions in Eq. (\ref{definition_lambda}) and introduced $\hat{\mathbf B}_{\chi} = \mathbf B_{\chi} / |\mathbf B_{\chi}|$.

\bibliography{Refs_WSM.bib}

\end{document}